\documentclass[]{interact}

\usepackage{epstopdf}% To incorporate .eps illustrations using PDFLaTeX, etc.
\usepackage[caption=false]{subfig}% Support for small, `sub' figures and tables
\usepackage[normalem]{ulem}
\usepackage[natbibapa,nodoi]{apacite}
\usepackage{xcolor}
\usepackage{tabularx}
\theoremstyle{plain}% Theorem-like structures provided by amsthm.sty

\theoremstyle{definition}

\theoremstyle{remark}

\usepackage[colorinlistoftodos]{todonotes}
\usepackage[autostyle]{csquotes}
\usepackage[british]{babel}

\begin{document}

% \articletype{ARTICLE TEMPLATE}% Specify the article type or omit as appropriate

\title{Prospects and applications of photonic neural networks}

\author{
\name{Chaoran Huang \textsuperscript{a}\thanks{CONTACT Bhavin J. Shastri. Email: bhavin.shastri@queensu.ca}, 
Volker J. Sorger\textsuperscript{b},
Mario Miscuglio\textsuperscript{b} ,
Mohammed Al-Qadasi\textsuperscript{c},
Avilash Mukherjee\textsuperscript{c},
Sudip Shekhar\textsuperscript{c},
Lukas Chrostowski\textsuperscript{c},
Lutz Lampe\textsuperscript{c},
Mitchell Nichols\textsuperscript{c},
Mable P. Fok\textsuperscript{d},
Daniel Brunner\textsuperscript{e},
Alexander N. Tait\textsuperscript{f},
Thomas Ferreira de Lima\textsuperscript{a},
Bicky A. Marquez\textsuperscript{g},
Paul R. Prucnal \textsuperscript{a},
Bhavin J. Shastri\textsuperscript{g}.}
\affil{\textsuperscript{a}Princeton University, Princeton, NJ 08542, USA; \\
\textsuperscript{b}The George Washington University, Washington DC, DC, USA;\\
\textsuperscript{c} The University of British Columbia, Vancouver, BC V6T 1Z4, Canada;\\
\textsuperscript{d} University of Georgia, Athens, GA 30602, USA;\\
\textsuperscript{d}  Institut FEMTO-ST, Universit\'e Bourgogne-Franche-Comt\'e CNRS UMR 6174, Besan¸con, France;\\
\textsuperscript{f} National Institute of Standards and Technology, 325 Broadway, Boulder, Colorado 80305, USA;\\
\textsuperscript{g} Queen’s University, Kingston, ON K7L 3N6, Canada}
}

\maketitle

\begin{abstract}
Neural networks have enabled applications in artificial intelligence through machine learning, and neuromorphic computing. Software implementations of neural networks on conventional computers that have separate memory and processor (and that operate sequentially) are limited in speed and energy efficiency. Neuromorphic engineering aims to build processors in which hardware mimics neurons and synapses in the brain for distributed and parallel processing. Neuromorphic engineering enabled by photonics (optical physics) can offer sub-nanosecond latencies and high bandwidth with low energies to extend the domain of artificial intelligence and neuromorphic computing applications to machine learning acceleration, nonlinear programming, intelligent signal processing, etc. Photonic neural networks have been demonstrated on integrated platforms and free-space optics depending on the class of applications being targeted. Here, we discuss the prospects and demonstrated applications of these photonic neural networks.

\end{abstract}

\begin{keywords}
Neuromorphic computing; photonic neural networks; neuromorphic photonics; silicon photonics; machine learning.
\end{keywords}

\section{Primer on artificial intelligence, machine learning, neuromorphic computing, and neuromorphic photonics}

Creating a machine that can process information like human brains has been a driving force of innovations throughout history. Artificial intelligence (AI) has been coined as an academic discipline since the 1950s~\citep{mccarthy2006proposal}. This field underwent the first surge of optimism from the 1950s to 1970s, however, followed by decades of setbacks. The biggest obstacle at that time was the lack of computing power~\citep{noauthor_history_2017}. In the last decade, AI has experienced explosive growth. Three sources fuel the advancement of AI: (1) substantial development of AI algorithms, especially in machine learning and neural network models (alias for “deep learning” \citep{lecun2015deep}); (2) the abundant amount of available information in the “big data” era; and (3) the rise of computing power as predicted by Moore’s Law, together with new hardware (e.g., graphics processing unit (GPU)) and infrastructures (e.g., cloud-based servers).

State-of-the-art AI algorithms, by and large, are implemented using neural networks, a computing model inspired by the brain's neuro-synaptic framework. Today, nearly all AI algorithms are running on digital computers based on von Neumann architecture, a computing architecture that has dominated computing design since it was invented but is nothing like the brain. This architecture consists of a centralized processing unit (CPU) that performs all operations specified by the program's instructions and a separate memory that stores data and instructions. It processes information sequentially in a serialized manner. However, neural network models are radically different from von Neumann architecture in some key features. First, neural networks are highly parallel and distributed, whereas von Neumann architecture is inherently sequential (or, in the best case: sequential-parallel with multi-processors). Second, in neural networks, computing units (neurons) and storage units (synapses) are co-located. In contrast, computing units (CPUs) and storage units (dynamic random access memories (DRAMs)) are physically separate chips in digital computers. The sharp contrast between the two architectures slows down the computing speed and increases the power consumption, which, as a result, necessitates reinventing conventional computers for efficient information processing.

Neuromorphic (i.e. neuron isomorphic) computing promises to solve these problems by creating radical new hardware platforms that can emulate the underlying neural structure of the brain. The general idea is to build circuits composed of physical devices that mimic the neuron biophysics interconnected by massive physical interconnects with co-integrated non-volatile memories. In doing so, neuromorphic hardware could break performance limitations inherent in von Neumann architectures and gain advantages in speed and efficiency in solving intellectual tasks. Achieving this goal requires significant advances in a wide range of technologies, including materials, devices, device fabrication, system integration, platform co-integration, packaging etc~\citep{schuman2017survey,berggren2020roadmap,shastri2021photonics}.

Neuromorphic hardware has been built in electronics on various platforms, including traditional digital CMOS \citep{merolla2014million,schemmel2010wafer,furber2014spinnaker} and hybrid CMOS-memristive technologies (discussed next) \citep{yang2013memristive,govoreanu201110}. Neural network models highlight the essential needs of high-degree physical interconnections, which, in electronic neuromorphic hardware, is achieved by incorporating a dense mesh of wires overlaying the semiconductor substrate as crossbar arrays. Unfortunately, electronic connections fundamentally suffer harsh trade-offs between bandwidth and interconnectivity \citep{nahmias2019photonic,miller2009device}. A major limitation for neuromorphic electronics is interconnect density, thus confining the neuromorphic processing speed and associated application space within the MHz regime.

Photonics has unmatched feats for interconnects and communications in terms of bandwidth, which can negate the bandwidth and interconnectivity trade-offs~\citep{shastri2021photonics,Ahm16,Ahm20}. The advantages of photonics for neural networks were recognized decades ago. The photonic neural network research was pioneered by Psaltis and others who adopted spatial multiplexing techniques enabling all-to-all interconnection \citep{psaltis1985optical}. However, low-level photonic integration and packaging technologies hindered the practical applications of photonic neural networks at that time. Nevertheless, the landscape of photonic neural networks has changed tremendously with the emergence of large-scale photonic fabrication and integration techniques \citep{shastri2021photonics,thomson2016roadmap,lin2018all,feldmann2019all}. For example, silicon photonics provides an unprecedented platform to produce large-scale and low-cost optical systems \citep{thomson2016roadmap,sun2013large,She21}. In parallel, a broad domain of emerging applications (such as solving nonlinear optimization problems or real-time processing of multichannel, gigahertz analog signals) is also looking for new computing platforms to fulfill their computing demands \citep{de2019machine,khan2019optical,ma2020blind,han2018solving,104huang2020demonstration}. All these changes have shed light on new opportunities and directions for photonic neural networks \citep{prucnal2017neuromorphic}.

This paper is intended to provide an intuitive understanding of photonic neural networks and why, where, and how photonic neural networks can play a unique role in enabling new domains of applications. First, we discuss the challenges of digital versus analog approaches in implementing neural networks. Next, we provide a rationale for photonic neural networks as a compelling alternative for neuromorphic computing compared to electronic platforms. Then, we outline the primary technology required for evolving neuromorphic photonic processors, review existing approaches, and discuss challenges. In the subsequent sections, we provide a survey of new applications enabled by photonic neural networks and highlight the role of photonic neural networks in addressing the challenges in these applications.

\section{Digital vs. analog neural networks }

In this Section, we briefly compare the state-of-the-art electronic implementations of neural networks in digital and analog domain. We establish the advantages and limitations of analog implementations in general to then make the case for analog photonic implementations in the following Sections.

Deep neural networks (DNN) model complex nonlinear functions by composing layers of linear matrix operations with non-linear activation functions. Computationally, DNNs are mostly matrix-multiplication, with matrix-multiplications taking more than 90\% of the total computations in a DNN \citep{cong2014minimizing}. Due to the underlying array-based operation in the matrix multiplication, digital electronic neural network hardware is usually composed of basic units, referred to as processing elements (PEs), in a 2D array structure \citep{chen2016eyeriss1}. Such a structure enables the matrix multiplication operation to be N$\times$ faster than CPUs, where N is the input vector length. Usually, PEs are composed of digital multipliers and adders, with precision up to 32 bits, to perform a single multiply and accumulate (MAC) operation, similar to the arithmetic logic unit in the CPU core.

Since DNNs consume a huge chunk of energy in data movement, many digital neural network hardware focus on optimizing dataflow to save energy. Based on the connection of PEs and the interconnects, various dataflows can be described. An output-stationary dataflow performs all the MAC operations for a single output before moving to the next. All the inputs and weights required are fetched from the memory, multiplied, and added to the partial sum, which is stored inside PE \citep{chen2016eyeriss}. On the other hand, a weight-stationary dataflow holds the weights inside PE to maximize weight reuse. The partial sum accumulation occurs across multiple PEs while the input vector is fed in a staggered style allowing the PEs to perform MAC operation with the internally stored weights. Further dataflow optimizations combining different types of data reuse are also possible to reduce energy further \citep{chen2016eyeriss,gudaparthi2019wire}.

Implementing the MAC operations in analog domain can help in reducing the energy consumption. Analog electronic elements, such as charge, current and time can be used to represent the data values. \textit{An inherent advantage in analog techniques is the built-in addition operation without requiring additional circuits.} To perform the MAC operation with analog electronics, switched-capacitor techniques charge a capacitance sized proportionally to the weight with a current sized proportionally to the input \citep{bankman20168}, current-steering techniques control the magnitude of current flowing through transistors \citep{skrzyniarz201624}, and time-domain techniques modulate the pulse width of a signal using controlled oscillators \citep{8820002}. Such analog techniques have shown to decrease energy significantly for small DNN models: 4$\times$ using switched-capacitor on BinaryNet, 67$\times$ using current steering on Matched Filter, and 1.4$\times$ using time-domain techniques on mobile reinforcement learning. The shortcomings of analog techniques include: limited size of DNN models, low bit precision ($<$ 4 b) and associated accuracy loss, analog-to-digital converter/digital-to-analog converter (ADC/DAC) overhead, susceptance to noise and process, voltage and temperature (PVT) variations. 

Analog implementations have a direct consequence for noise and noise propagation \citep{doi:10.1162/089976698300017052}. Since the probability of corrupting a symbol usually is identical for all bits in a sequence, the impact of a noise-induced Boolean symbol modification can be dramatic. Compared to that the corruption of an analog signal is usually more subtle as signal perturbations are mostly proportional to noise amplitude. Digital encoding therefore requires that thresholding levels significantly exceed all noise amplitudes; however, the signal propagation is then practically noiseless as the noiseless symbolic representation is continuously re-established. Furthermore, increasing a digital signal’s resolution is comparatively economic as the number of digitization levels grows exponentially.

Well-designed circuits readily approach the thermodynamic noise limit to better than one order of magnitude \citep{4358095}. Thermodynamics, therefore, establishes the link between such information centered arguments and the energy fundamentally required for a certain SNR. Digital encoding is penalized with a large constant energy penalty due to the required high encoding fidelity. However, its superior scaling means that digital becomes more energy efficient than analog, roughly beyond SNR $<10^4$ \citep{7878958}. Recently digital implementations of neural networks significantly reduced their bit-precision, with several systems today running with 8 or less bit resolution during inference. Finally, spiking NNs occupy a middle ground and potentially are superior in energy efficiency to analog for SNR $>10^2$ and to digital for SNR $<10^7$ \citep{7878958}. 

Finally, the accumulation of noise can be strongly managed using the connections of a neural network. Studies based on linear, symmetric, i.e. untrained networks of noisy linear neurons show that neural network analog in and output neurons are the chief noise source, while in particular noise uncorrelated across neurons is essentially fully suppressed through the network's connections \citep{doi:10.1063/1.5120824}. New studies in fully trained networks of noisy nonlinear units show that nonlinearity also efficiently decorrelates noise from correlated noise-populations \citep{noiseintrain}. This is important as such noise can for example be induced by a common power supply. Finally, rather weak requirements allow to fully freeze the propagation of noise through a network, and an analog photonic neural network’s output can, therefore, approach the SNR of a single neuron \citep{noiseintrain}.

Another possible method to reduce data movement energy is moving the computing inside the memory modules itself. Such architectures are referred to as In-Memory computing (IMC), and use the memory cells as an analog circuit to perform the MAC operations, generally in a weight-stationary dataflow. The inputs are analog currents or voltages on the wordlines, while their weights are either binary, ternary or digitally stored over multiple memory cells. The accumulation happens inherently in the bitlines in the memory array, resulting in an analog output \citep{7573556,8310397}. IMC architectures, while significantly enhancing the throughput of the system, operate in analog which call for adding more system level design considerations to meet the output signal to noise ratio (SNR) requirements and size.

IMC can also be performed in crossbar arrays of emerging memories, such as resistive RAM (ReRAM), conductive bridging, magnetic tunnel junctions, and phase change memories \citep{9075887}. Explicit multipliers, adders and PE interconnects are not needed in IMC. Rather, the equivalent PE array in digital is implemented in IMC in just the area required to implement the memory array, plus the ADC and DAC at the array periphery \citep{9075887}. Increased area efficiency allows packing far more parallel units, hence processing operations can be accelerated by more than an order of magnitude. For comparison, the area required to implement the PE in digital implementations is $13.4\times10^6$ F\textsuperscript{2}  \citep{chen2016eyeriss1}, whereas the average memory cell sizes in IMC are less than 100 F\textsuperscript{2} \citep{9062953}, where F is the minimum feature size of the technology. Furthermore, energy consumption can also be reduced by an order of magnitude over the equivalent digital systems, since several MAC operations are performed almost at the cost of a single read operation of the memory array. 

IMC suffers from constraints similar to the analog MAC implementations. Furthermore, the nonidealities of analog memory cells and their interconnects pose limitations for achieving high accuracy and scaling \citep{8267068}. For example, the nonlinearities of memory cells and the resistance of the interconnect in ReRAM IMC was shown to degrade the accuracy of computation with scaling \citep{7058989}. Noise limitations have shown to saturate the computing accuracy of analog crossbars to 8 bits \citep{7544263}. In comparison with digital implementations, the energy, area and latency advantages from IMC have been shown to reduce with the increased precision \citep{8804680}. But precision reduction techniques such as variable layer precisions and nonuniform quantization \citep{cong2014minimizing,7783722}, have shown equivalent accuracy to 32-bit digital implementations even after reducing precision down to 2-bit \citep{choi2019accurate}. 

%I have kept this section limited to electronics only, no discussion on photonics. 
% \sout{These are exceptional arguments in favor of photonic neural networks, as digital photonic computing has essentially been abandoned due to the large energy penalty in binary switching, and since analog photonic systems \sout{readily}\textcolor{red}{can} achieve SNRs comparable to 6-8 bit precision \citep{AndreoliPorte,doi:10.1063/1.5120824}. }

\section{The case for photonics for neuromorphic processors} % (fold)
\label{sec:the_case_for_photonics_for_neuromorphic_processors}

In an artificial neural network, neurons are interconnected by synaptic weights (a memory element). Signals from many neurons are weighted before being summed by the receiving neuron. This many-to-one (N:1) connection is called fan-in, and the weighted sum–a linear operation–is the dot product of the output from connected neurons attenuated by a weight vector. A neuron then performs a nonlinear operation on the weighted sum to implement a thresholding effect which is output to many neurons. This one-to-many (1:N) connection is called fan-out.

Electronic and photonic neuromorphic approaches to implement neural networks face different challenges. Physical laws that restrain electronics do not necessarily apply to optics. For high speed data transfer, compared to electronic wires, optical waveguides have a lower attenuation, have no inductance (minimal frequency-dependent signal distortions), and photons hardly interact with other photons (unless intermediated by matter) which allows for wavelength multiplexing. While these fundamental properties have proven to be advantageous in optical communications, they are also important for neural networks including interconnectivity \citep{1457211,867687} (i.e. neuron-to-neuron, neurons-to-neurons and neurons-to-memory communications), and parallel and linear processing i.e. matrix multiplication \citep{867687}. However, these same properties make it challenging to implement nonlinear operations \citep{doi:10.1080/713821757}. We refer the reader to the quantitative analyses comparing electronic and optical interconnects, and electronic and photonic computing performed by Miller \citep{miller2009device,867687} and Nahmias et al \citep{nahmias2019photonic,FerreiradeLimaTaitMehrabianNahmiasHuangPengMarquezMiscuglioElGhazawiSorgerShastriPrucnal+2020+4055+4073}, respectively. Here, we qualitatively summarize these concepts comparing and contrasting optics and electronics.

\textbf{Interconnects}: Conventional electronic processors rely on point-to-point memory processor communication and can take advantage of state-of-the-art transmission line and active buffering techniques \citep{shastri2021photonics}. However, a neuromorphic processor typically requires a large number of interconnects (i.e., hundreds of many-to-one fan-in per processor) \citep{10.3389/fnins.2013.00118} where it is not practical to use line and active buffering techniques for each connection at high speed. This creates a communication burden which in turn, introduces fundamental performance challenges that result from RC and radiative physics in electronic links, in addition to the typical bandwidth-distance-energy limits of point-to-point connections \citep{867687}. While some electronic neuromorphic processors incorporate a dense mesh of wires overlaying the semiconductor substrate as crossbar arrays, large-scale systems employ time-multiplexing or packet switching, notably, address-event representation (AER). These virtual interconnectivities allow electronic approaches to exceed wire density by a factor related to the sacrificed bandwidth. As argued by \cite{shastri2021photonics}, for many applications, however, bandwidth and low latency are paramount, and these applications can be met only by direct, non-digital photonic broadcast (that is, many-to-many) interconnects.

Apart from bandwidth, power dissipation in interconnects is another major challenge in neuromorphic processors. Data movement has posed severe challenges to the power consumption of today's digital computers as parallel computing is widely deployed. A large amount of data has to move among processors and memory units, especially in those distributed programming models like neural networks. In these models, most energy is lost in data movement \citep{10.1145/3140659.3080246} due to capacitive charge-discharge events at high frequencies. In contrast, optics is more energy efficient in data movement at high speeds.  

% \sout{When data propagates in the metal wires, the wires are constantly charged and discharged. The power dissipation caused by data movement is proportional to the wire capacitance, which increases with the wire length.} In contrast, optics \sout{has superior performance} \textcolor{red}{is more energy efficient} in data movement \textcolor{red}{at high speeds}. \sout{The propagation loss of optical waveguides can be designed to be very small. For example, a typical silicon waveguide loss is around 1 dB/cm \citep{10.1117/12.2076262}. Therefore, within the size of an integrated chip, communication caused power dissipation is almost zero, which contrasts sharply with electronics.} 

\textbf{Linear operations and weighting}: Optical physics is very well suited for linear operations. The electric field or intensity of the light can also be used as an analog metric for neuromorphic computations. The analog input in the electrical domain (voltage) can be used to modulate components such as Mach Zehnder (MZ) modulators (MZMs) or ring resonator (RR) modulators (RRMs). The weights can be implemented by scaling the field with MZ interferometers (MZIs) \citep{shen2017deep} or the intensity with RRs \citep{tait2014broadcast,Tait18}. Optical signals can be added through wavelength-division multiplexing (WDM) by accumulation of carriers in semiconductors \citep{tait2014broadcast}, electronic currents \citep{shainline2017superconducting, bangari2019digital} or changes in the crystal structure of a material induced by photons \citep{feldmann2019all}. Larger vector matrix multiplications (VMM) can thus be implemented with such MZ \citep{shen2017deep} or RR \citep{tait2014broadcast} based circuits. Such photonic neural networks, being analog in nature, share the advantages and constraints of their analog electronic counterparts. In addition, modulations can be done at 10s of GHz, addition can be done at the speed of light in coherent MZI networks and at 10s of GHz in RR networks, and weight multiplication can be done at the speed of light, all of which provides much lower latency than analog or digital electronic MAC operations. 

Dynamic power consumption is also much lower, since multiplication does not consume any energy and energy efficiency of the E/O modulation can be reduced to as low as 10 fJ per operation \citep{nozaki2019femtofarad}. However, static power consumption is dependent on laser wall plug efficiency, waveguide losses, energy consumed in maintaining the weights, etc, and must be minimized.  Assuming an estimated energy consumption of $>$ 100 fJ per operation for digital implementation \citep{lin20207}, analog IMC brings 2.1-7.8$\times$ reduction \citep{xue201924,9062953}. Photonic implementations can achieve energy consumption similar to IMC, but with orders of magnitude reduction in latency. Like their IMC counterparts, photonic neural networks can operate up to 8 bits of precision \citep{ramey2020silicon,bangari2019digital,huang2020demonstration}.  

\textbf{Nonlinear operations}: The same properties that allow optoelectronic components to excel at linear operations and interconnectivity are at odds with the requirements of nonlinear operations for computing \citep{shastri2021photonics}. The implementation of photonic neurons relies on the nonlinear response of optical devices. The approaches fall into two major categories based on the physical representation of signals within the neuron: optical-electrical-optical (O/E/O) vs. all-optical. 

O/E/O neurons involve the transduction of optical power into electrical current and back within the signal pathway. Their nonlinearities occur in the electronic domain and in the E/O conversion stage using lasers or saturated modulators  \citep{tait2019silicon,80amin2019ito,81george2019neuromorphic}. In the authors' recent work \citep{tait2019silicon}, neurons are implemented using silicon modulators that exploit the nonlinearity of the electro-optic transfer function. Modulation mechanisms can change the real and imaginary parts of the refractive index of the material and subsequently alters the index (termed EO modulators) and loss (termed electro-absorptive modulators (EAMs)) of the optical propagating mode. Research on high speed and power-efficient modulators are very active, with a general focus on maximizing the interaction between the active material and the light. Such approaches include lithium niobate (LiNbO3) modulators based on the Pockels effect \citep{wang2018integrated}, III-V semiconductor-based quantum-confined Stark effect modulators \citep{kuo2005strong}, silicon modulators based on plasma dispersion effect \citep{xu2005micrometre,dong2009low}, and hybrid modulators incorporating novel materials (such as ITO and graphene) to silicon-based modulators \citep{80amin2019ito,liu2011graphene,komljenovic2016heterogeneous,amin20180}. With the foundry-compatible silicon-on-insulator (SOI) technology for on-chip integrated photonics, OEO neurons are demonstrated by \cite{tait2019silicon} using a silicon microring resonator (MRR) with embedded PN modulator, and \cite{williamson2019reprogrammable} with a Mach-Zehnder type modulator.

All-optical neurons rely on the semiconductor carriers or optical susceptibility that occur in many materials. A perceived advantage of all-optical neuron implementations is that they are inherently faster than O/E/O approaches due to relatively slow carrier drift and/or current flow stages in the latter. All-optical perceptron has been demonstrated based on single-carrier optical nonlinearities, including through carrier effect in MRR \citep{huang2019programmable,jha2020reconfigurable}, changing a material state \citep{feldmann2019all,chakraborty2018toward}, such as via a structural phase transition, and saturable absorbers and quantum assemblies heterogeneously integrated in PICs \citep{79miscuglio2018all}.

\textbf{Memory}: As dicussed in Section 2, nonvolatile materials allow implementing memory elements directly on the computing devices for IMC. On-chip nonvolatile memories that can be written, erased, and accessed optically are rapidly bridging a gap toward on-chip photonic computing \citep{rios2019memory}; however, they cannot usually be written to and read from at high frequencies. As described by \cite{shastri2021photonics}, future scalable neuromorphic photonic processors will need to have a tight co-integration of electronics with potentially hybrid electronic and optical memory architectures, and take advantage of the memory type (volatile versus non-volatile) in either digital or analog domains depending on the application and the computation been performed.

Current approaches with photonic neural networks are driven by electronic circuits or micro-controllers to load matrices. The integration and packaging of large-scale optical and electronic circuits can be challenging in terms of cost and power. In some machine learning applications (such as deep learning inference) the weights, once trained, do not have to be updated often or at all. In these cases, the integration of novel photonic memory technologies can limit the need to read from and write to electronic memories with DACs and ADCs. All-optical memories have been demonstrated using various optical components such as nonlinear switches \cite{95dorren2003nonlinear}, MZIs \citep{96hill2001fast}, laser diodes \citep{97hill2001all}, semiconductor optical amplifiers (SOAs), bandpass filters (BPFs) and isolators \citep{98wang2008soa}. The access time of an optical memory cell is small and attractive for photonic neural networks \citep{99alexoudi2020optical}. Non-volatile photonic memories with phase-change materials (PCMs) set and retain the weights without further holding power after being set \citep{feldmann2019all}, resulting in almost zero power consumption in performing matrix multiplication operations. Here a crystalline phase transition (from amorphous to crystalline and back) constitutes reversible (WRITE $\&$ RESET) memory programmability, therefore enabling dynamic synaptic plasticity and online learning in photonic neural networks. For online training of the photonic neural network, a high-speed WRITE would be ideal, and experimentally MHz speeds are possible for known PCMs such as GST or GST alloys, just limited by the heat-capacitance of these photonic RAM. The READ speed, interestingly, is on the order of ps, and simply given by the time-of-flight of the signal photon through such photonic RAM. However,significant improvements must be made in reducing the energy consumption (taking into account optical losses) and size, reliability and ease of interfacing for their practical deployment in photonic neural networks. The current prominent material GST has prohibitively high optical losses even for the lower-loss amorphous state with an extinction coefficient (kappa = 0.2), thus leading to high insertion loss. Future research should focus on low-optical loss solutions. Emerging P-RAMs would also eliminate the long-standing \lq memory access bottleneck\rq~known from electronics, and a successful P-RAM integration constitutes a similar shift in optics from centralized to decentralized computing, known as in-memory computing, one of the very active research fields left in circuit design, after transistor scaling stopped.

% While there is no photon-photon force in the universe, a photonic memory as such, can only be realized using matter (i.e. light-matter-interactions). Fortunately, such state-retention is recently achieved in electro-thermal programmable PCM; here a crystalline phase transition (from amorphous to crystalline and back) constitutes reversible (WRITE $\&$ RESET) memory programmability. For online training of the photonic neural network, a high-speed would be ideal, and experimentally MHz speeds are possible for known PCMs such as GST or GST alloys, just limited by the heat-capacitance of these photonic RAM . This READ speed, interestingly, is on the order of ps, and simply given by the time-of-flight of the signal photon through such photonic RAM. However, the current prominent material GST has prohibitively high optical losses even for the lower-loss amorphous state with an extinction coefficient (kappa = 0.2), thus leading to high insertion loss. Future research should focus on low-optical loss solutions. Indeed, such emerging P-RAMs would also eliminate the long-standing \lq memory access bottleneck\rq~known from electronics, and a successful P-RAM integration constitutes a similar shift in optics from centralized to decentralized computing, known as in-memory computing, one of the very active research fields left in circuit design, after transistor scaling stopped. 

\section{Architectures of neuromorphic photonic processors} % (fold)
\label{sec:architectures_of_neuromorphic_photonic_processors}

\begin{figure}[h!]
\centering
\includegraphics[width=0.75\linewidth]{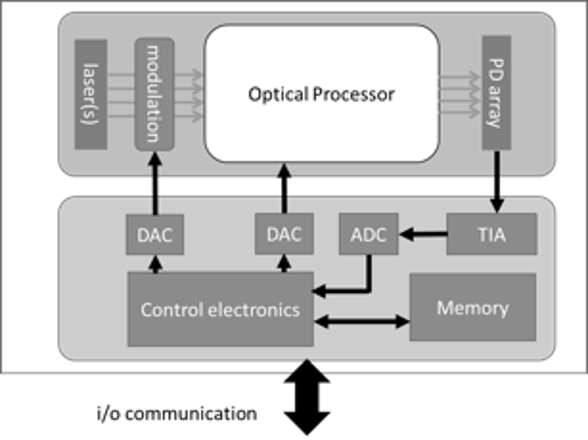}
% \vspace{-10pt}
\caption{Generic system schematic of an analog photonic neural network or photonic tensor core accelerator. The digital-to-analog domain crossings are power costly and ideally a photonic DAC without leaving the optical domain, e.g. Ref \citep{meng2019electronic} would be used instead, thus reducing system complexity and hence allowing for extended scaling laws. The optical processor itself performs the MAC operation such as for VMM, convolutions, or of a perceptron, but could also include nonlinearity (neuron thresholding) for higher biological plausibility (e.g. spiking neurons where event-driven scheme play a role)\citep{tait2019silicon,nahmias2020laser,jha2020reconfigurable,peng2018neuromorphic}. }
\label{fig:figure1}
\end{figure}

Neuromorphic processors and photonic neural networks alike require a number of fundamental building blocks; i) synaptic MAC operation, ii) nonlinear activation function (NLAF); iii) state-retention, i.e. memory, iv) data input/output (I/O) ports and, depending on the applications, data domain crossings such as v) photonic to electronic and/or vi) analog-to-digital, with the latter including DAC and the ADC counterparts. However, a generic architecture of a current state-of-art photonic neural network system is given in Figure.~\ref{fig:figure1}

\textbf{Photonic synaptic MAC operations and VMMs}: The most common form of photonic neural networks focuses on accelerating the mathematical computation of multiplications using optics; this is not surprising, since photonic programmable circuits allow for a non-iterative mathematical multiplication by simple preparing the state of the programmable element followed by sending the optical signal through this element. Then, the multiplication is performed `on-the-fly' at pico-second delays in PICs enabling a notion of `real-time' (i.e. zero-delay) computing, which is incidentally not to be confused with computing schemes that `expect' a system to complete a task at a pre-determined and deterministic time. Options for VMM are plenty, but the most common ones are a mesh network of cascaded MZIs, or MRR filters, or a photonic tensor core (PTC) processor~\citep{shen2017deep,tait2017neuromorphic,miscuglio2020massively}. Accelerating VMMs is a worthy aim, since the mathematical computational complexity of VMMs scales with N\textsuperscript{3}, where N is the matrix size (assuming a square matrix). However, the mathematical complexity is technically speaking irrelevant (at first order), since for hardware implementations considered here, the runtime complexity is actually of interest. And it is this runtime complexity that is non-iterative in analog photonic neural networks (assuming the neuron’s weights are set, i.e. programmed). 

\textbf{NLAF, or threshold and training neural networks}: Depending on the underlying algorithm model of the neural network, the NLAF can be as simple as a step-function such as used in binary classification problems, or a more elaborate function such as an sigmoidal, tangent hyperbolic, or population growth functions. However, when performing neural network training using gradient descent (GD) backpropagation (BP) methods, the problem of vanishing (or exploding) gradients can occur, where during each training cycle (called epoch) the available gradient on which a differentiation is performed is becoming consecutively smaller to the point of noise-level dominated and training would seize. To prevent this, rectifying linear units (ReLU) or Gaussian error linear units (GELU) could be used instead. In a ReLU, for example, the output is ‘zero’ up until a certain input level, upon which, the output is simply a linear function. This linearity ensures a constant and non-zero gradient during GD BG training cycles. Since differentiation at a step is mathematically not defined, a Soft-ReLU is often used instead to ensure continued loss-function minimization and neural network performance improvement with training. Note, the known problem of overfitting neural networks does also apply to photonic neural networks. However, pruning techniques and hyperparameter adjustments during training are an known and interesting method (yet time consuming) to improve artificial neural network performance. Interestingly, for the context of photonic neural networks, the network interconnectivity sparseness created in pruning steps, saves fan-out (fan-in) connections as well as waveguide connections. This is a blessing, since the bulkiness of photonic components (in contrast to electronic counterparts on a per-unit basis) increases functional density and thus performance per unit chip footprint. Indeed, future work should explore pruning techniques further for photonic neural networks. An initial study was, for instance, performed on the MZI NxN mesh network that was originally developed by the MIT groups ~\citep{shen2017deep}, where it was shown that the number of required MZIs can be reduced from a brute-force N$\times$N \citep{78gu2020efficient}. Incidentally, in performance-optimized DNNs, the specific shape of the NLAF should be adjusted with layer depth; for instance, a ReLU that is not saturating is more useful in the upstream layers, while a saturating sigmoidal function supports a \lq decision-making\rq~process at the fully-connected (FC) layer at the output in classification problems. In most neural networks demonstrations, the primary performance driver is power efficiency, while speed is a secondary consideration. Therefore, early photonic neural network demonstrations implement NLAF in the digital electronic domain, limiting the speed of neural network to the clock speed (hundreds of MHz to a few GHz). Nevertheless, optical NLAF, as discussed in Section~\ref{sec:the_case_for_photonics_for_neuromorphic_processors} offer a unmatched speed over ten GHz in performing NLAF, thus becoming essential elements in order to solve many compelling applications that requires online (i.e., real-time) learning and inference or for neural networks with gigahertz bandwidths.

% do not consider optical nonlinearity Optical NLAF, as discussed in Section~\ref{sec:the_case_for_photonics_for_neuromorphic_processors}, can 
% As discussed in Section~\ref{sec:the_case_for_photonics_for_neuromorphic_processors}

% \textcolor{red}{THE REST OF THIS PARAGRAPH CAN BE DELETED or SIGNIFICANTLY TRIMMED, since optical nonlinearity was discussed already in Section 3.}Going back to photonics, early photonic neural network demonstrations and most photonic neural networks at the time of writing this paper do not consider optical nonlinearity. The Sorger group explored in an earlier paper some emerging concepts for all-optical NLAF using saturable absorbers and quantum assemblies heterogeneously integrated in PICs \citep{79miscuglio2018all}. An alternative option is to perform nonlinearity electro-optically such as using EO modulators \citep{80amin2019ito,81george2019neuromorphic,tait2019silicon}, switches, or any other form of programmable photonic element. Electro-optic NLAF do offer synergies for photonic neural networks and photonic tensor cores (PTC) where the summation for the MAC operation is performed non-coherently using photodetectors, and the weighted sum of the input signals is already converted from the optical VMM into an electrical signal by the detector, and its signal can be fed to the EO modulator as a gate at the photonic neuron’s output.  

\textbf{Memories}: The memory is usually implemented electronically and various choices exist. Static RAM (SRAM) and dynamic RAM (DRAM) are used in neural network architectures to store inputs, weights, training parameters and look up table (LUT) values. At the cell level, SRAM typically uses 6 transistors (6T) to store a single bit, whereas DRAMs use a single transistor and a single capacitor (1T1C). SRAMs are indispensable for neural network implementations given their faster access time; in addition, they can be leveraged for data reuse to reduce data fetch energy \citep{89sze2017efficient}. DRAM with its larger storage capacity is used to store all activations and training weights. However, owing to off-chip implementation, DRAM is slower with higher energy consumption than SRAM. Depending on the size of an SRAM and its location, the energy for read and write can scale from sub-pJ to 10 pJ per byte for SRAMs \citep{gudaparthi2019wire,90horowitz20141}. With an estimated SRAM size of 64KB used for DNN, the corresponding read and write latencies are $\sim$1 ns \citep{91imani2016low}, which can be projected to reduce to 0.25 ns in 7-nm CMOS. Given the size difference and off-chip implementation in comparison to SRAMs, off-chip DRAMs consume more than two orders of magnitude of energy for data fetch \citep{89sze2017efficient}. High bandwidth memories (HBM) are used to reduce the energy consumption of DRAM's data fetch \citep{92tran2016era}, by moving DRAM modules closer to the chip. Alternatives to DRAM include resistive memories and PCMs. Utilizing 1T1R in ReRAM crossbar topology was demonstrated with low energy consumption and write latency \citep{93xiao2020analog}. On the other hand, using PCMs as unit cells requires high energy to write, and has poor resolution and linearity \citep{94Mukherjee2021case}. 

An analog memory cell made up of a capacitor has drawbacks such as need for ADC/DACs to interact with digital cells, low density compared to SRAM cell, need for refresh, and sensitivity to noise and crosstalk. However, for photonic computing, placing a capacitor memory cell next to photonic elements is attractive. This is because the computing is analog in nature, and photonic compute elements are much bigger than digital compute elements, so the large size of the capacitor memory cell in comparison to SRAM is not a major concern. Such an architecture can eliminate the data movement bottleneck, significantly reducing the access time and energy consumption. Monolithic photonic processes supporting metal capacitors are ideally suited for such implementations. 

For most applications, training the neural network is time consuming, spanning hours to days or even weeks. Thus, it is realistic to assume that for most applications, the neurons’ weights or the kernel of a PTC is fixed and does not change often in time. For this reason, a non-volatile solution that retains information-of-state (i.e. memory), is of high interest. If achieved, a (near) zero static power consumption can be achieved in photonic neural networks, allowing them to be rather efficient. Note, ‘static’ refers here to the MAC operation and not to possible signal modulation, which would be considered part of the I/O of the system. Fortunately, such state-retention is recently achieved in electro-thermal programmable PCM.

\textbf{Data I/Os}: Photonic accelerators such as photonic neural networks and PTCs allow for high-throughputs approaching P-OPS (peta operations per second). Such a photonic ‘highway’, while promising, may not demonstrate its full potential, if the to-be-processed input data is not provided at a sufficiently high data rate to the optical accelerator. This can be assured in two ways; either the I/O data bandwidth is sufficiently high such as provided by a FPGA or the data is already prevailing in the optical domain (such as of an optical aperture from a camera system, for example). The latter is elegant, since it not only eliminates the needs to drive power-costly EO modulators for signal encoding, but more importantly eliminates the requirements for DACs/ADCs (see next paragraph). Indeed, high-speed DACs would consume about a third of the total photonic neural network system’s power. For the case of optical data as the input, some PTCs show a dramatic power drop from about 80W down to 2W when DACs are not needed~\citep{82miscuglio2020photonic}. 

\textbf{Domain crossings: digital/analog domain crossings}: DACs and ADCs are required to interface a photonic neural network with digital signal processing (DSP) units (typically back-end) or when receiving data input data digitally such as from a server/computer etc. High sampling rate DACs used to drive input modulators mostly utilize current steering schemes and dissipate $>$5.5 pJ of energy for 6b-8b of conversion resolution \citep{Sedighi2011,kim2010}. The contribution of such converters to the overall energy efficiency of the neural network is reduced by 1/N as the network is scaled with N. Given the reusability of the weights over a given batch size, low speed capacitive DACs can be used. These DACs are usually adopted in SAR architectures and typically contribute to most of the ADCs' energy. Charge average switching, merge and split, charge recycling, and common mode voltage (Vcm) based charge recovery are some of the techniques used to reduce the DAC capacitance and consequently its switching energy \citep{Liou2013,Lin2015,Zhu2010}. In the Vcm-based scheme, the differential DAC arrays are connected to a common mode voltage Vcm which reduces the DAC’s switching energy. The power consumption of high speed ADCs scales exponentially with the resolution and linearly with the conversion rate \citep{Murmann2021}. For photonic neural network applications where the sampling frequency of the analog frontend is between 1-10 GS/s, and the resolution requirement is $<$ 8b, the energy consumption can be estimated as  $\sim$1 pJ for state of the art ADCs \citep{Murmann2021}. The architecture for ADC depends upon the photonic neural network. In recurrent neural network and long short-term memory (LSTM) networks, or in training back-propagation, the ADC is in a feedback network.  Hence, a low-latency ADC architecture must be chosen. The lowest achievable latency is achieved in Flash ADCs, approximately $\sim$100 ps (with an estimated delay of 80 ps for dynamic comparators and 30 ps for the encoding gates). Therefore, latency-dependent photonic neural networks are limited to $\sim$10 GS/s of operation. However, high-speed Flash ADCs also consume large power. On the other hand, neural network architectures such as convolutional neural networks which do not rely on feedback relax the low latency constraints for the ADC. In such implementations, pipeline and time interleaved SAR ADCs are better suited given their low energy consumption and high conversion rate.

\section{Applications of photonic neural networks} % (fold)
\label{sec:section_name}

Integrated optical neural networks will be smaller (hundreds of neurons) than electronic implementations (tens of millions of neurons). But the bandwidth and interconnect density in optics is significantly superior to that in electronics. This raises a question: what are the applications where sub-nanosecond latencies and energy efficiency trump the sheer size of processor? These may include applications 1) where the same task needs to be done over and over and needs to be done quickly; 2) where the signals to be processed are already in the analog domain (optical, wireless); and 3) where the same hardware can be used in a reconfigurable way. This section will discuss potential applications of photonic neural networks in computing, communication, and signal processing.

\subsection{High-speed and low-latency signal processing for fiber optical communications and wireless communications}

\subsubsection{Fiber nonlinearity compensation}

The world is witnessing an explosion of internet traffic. The global internet traffic has reached 5.3 exabytes per day in 2020 and will continue doubling approximately every 18 months. Innovations in fiber communication technologies are required to sustain the long-term exponential growth of data traffic \citep{100Cisco}. Increasing data rate and system crosstalk has imposed significant challenges on the DSP chips in terms of analog-to-digital converters (ADCs) performances, circuit complexity, and power consumption. A key to advancing the deployment of DSP relies on the consistent improvement in CMOS technology \citep{101pillai2014end}. However, the exponential hardware scaling of ASIC based DSP chips, which is embodied in Moore’s law as other digital electronic hardware, is fundamentally unsustainable. In parallel, many efforts are focused on developing new DSP algorithms to minimize computational complexity, but usually at the expense of reducing transmission link performances \citep{102agrell2016roadmap}. 

Instead of embracing such a complexity-performance trade-off, an alternative approach is to explore new hardware platforms that intrinsically offer high bandwidth, speed, and low power consumption \citep{de2019machine,103argyris2018photonic,104huang2020demonstration}. Machine learning algorithms, especially neural networks, have been found effective in performing many functions in optical networks, including dispersion and nonlinearity impairments compensation, channel equalization, optical performance monitoring, traffic prediction, etc \citep{khan2019optical}.

PNNs are well suited for optical communications because the optical signals are processed directly in the optical domain. This innovation avoids prohibitive energy consumption overhead and speed reduction in ADCs, especially in data center applications. In parallel, many PNN approaches are inspired by optical communication systems, making PNNs naturally suitable for processing optical communication signals. For example, we proposed synaptic weights and neuron networking architecture based on the concept of WDM to enable fan-in and weighted addition \citep{tait2014broadcast}. This architecture can provide a seamless interface between PNNs and WDM systems, which can be applied as a front-end processor to address inter-wavelength or inter-mode crosstalks problems that DSP usually lacks the bandwidth or computing power to process (e.g., fiber nonlinearity compensation in WDM systems). Moreover, PNNs combine high-quality waveguides and photonic devices that have been initially developed for telecommunications.  Therefore, PNNs, by default, can support fiber optic communication rates and enables real-time processing. For example, The a scalable silicon PNN proposed by the authors is composed of microring resonator (MRR) banks for synaptic weighting and O/E/O neurons to produce standard machine learning activation functions. The MRR weight bank is inspired by WDM filters, and the O/E/O neurons use typical silicon photodetector and modulator. Therefore, the optimization of associated devices in PNNs can utilize the fruits of the entire silicon photonic ecosystem that is paramountly driven by telecommunications and data center applications. 

In order to truly demonstrate photonics can excel over DSP, careful considerations are required to identify different application scenarios (i.e., long-haul, short-reach) and system requirements (i.e., performances, energy). Continuous research is needed to improve photonic hardware and to develop hardware-compatible algorithms. Here, we discuss several approaches to train and apply PNNs for optical communications.

% Photonic neural networks provide a promising hardware platform for optical communication signal processing. photonic neural networks can process optical communication signals in the optical domain, thus significantly relaxing the stringent requirements on ADCs circuits. Besides, photonic neural networks, which are composed of devices originally developed for telecommunications, are naturally suitable for processing high-speed optical communication signals in real-time and with low latency.

\begin{figure}[h!]
\centering
\includegraphics[width=\linewidth]{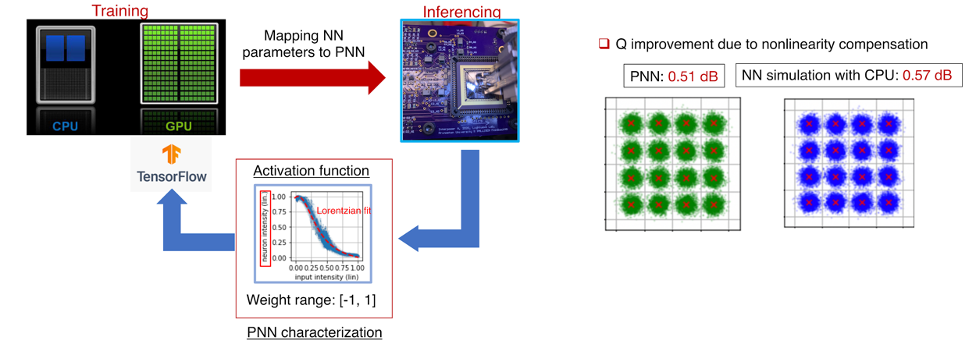}
% \vspace{-10pt}
\caption{(left) Concept of training and implementing photonic neural networks. Inset shows a transfer function of the photonic neural network measured with real-time signals. (right) Constellations of X-polarization of a 32 Gbaud PM-16QAM, with the ANN-NLC gain of 0.57 dB in Q-factor and with the PNN-NLC gain of 0.51 dB in Q-factor. \cite{104huang2020demonstration}}
\label{fig:figure2}
\end{figure}

\textbf{Neuromorphic approach} Long-haul communication systems prioritize high performances in terms of distance reach and spectral efficiency.  This requirement allows the use of coherent technology, along with dense wavelength multiplexing and polarization multiplexing schemes, to maximize the fiber capacity. In long-haul fiber optic transmission systems, fiber nonlinearity remains a challenge to the achievable capacity and transmission distance. One reason is that the nonlinear interplay between signal, noises, and optical fibers negates the accuracy of conventional nonlinear compensation algorithms based on digital backpropagation. Another reason is, the implementation of most nonlinear compensation algorithms in DSP chips demands excessive resources. In contrast, the neural network approach can learn and approximate the nonlinear perturbation from the abundant training data, rather than solely relying on the physical fiber model (known as stochastic nonlinear Schrodinger equation).  Based on the perturbation methods, the derived neural network algorithm has enabled compensating the nonlinear distortion in a 10800 km fiber transmission link with 32 Gbaud signals \citep{105zhang2019field}. \cite{tait2017neuromorphic} developed a photonic neural network platform based on the so-called "neuromorphic" approach, aiming to map physical models of optoelectronic systems to abstract models of neural networks (which differs from the reservoir approaches discussed next). By doing so, the photonic neural network system can leverage existing machine learning algorithms (i.e., backpropagation) and map training results from simulations to heterogeneous photonic hardware. The concept is shown in Figure~\ref{fig:figure2}.  A proof-of-concept experiment demonstrates the real-time implementation of a trained neural network model using an integrated silicon photonic neural network chip \citep{104huang2020demonstration}. In this work, the authors experimentally demonstarted that the silicon photonic neural network can produce a similar Q factor improvement compared to the simulated neural network for nonlinear compensation as shown in Figure~\ref{fig:figure2}, but it promises to process the communication data in real-time and with high bandwidth and low latency.

We also proposed a photonic architecture enabling all-to-all continuous-time recurrent neural networks (RNN) \citep{tait2017neuromorphic}. Recurrent neural networks can resemble optical fiber transmission systems: the linear neuron-to-neuron connections with internal feedback is analog to linear multiple-input multiple-output (MIMO) fiber channel with dispersive memory. With neuron nonlinearity, RNNs can be ideally used to approximate all types of linear and nonlinear effects in a fiber transmission system and compensate for different transmission impairments. RNNs, consisting of many feedback connections, are considered to be computationally expensive for digital hardware and require at least milliseconds to conduct a single inference. Contrarily, in photonic RNN, the feedback operations are simply done by busing the signals on photonic waveguides, allowing photonic hardware capable of converging to the solution within microseconds. This architecture also adopts the neuromorphic approach and thus allows to train PNNs externally using standard machine learning algorithms.

\subsubsection{Channel and/or predistortion equalization}

\textbf{Reservoir computing} Short-reach fiber-optic communication systems (FOCS) have recently seen increasing demand driven largely by the proliferation of cloud-based computing architectures and fronthauling in cloud radio access networks (C-RAN). DSP-based coherent transceivers are optimized for reach and capacity and generally considered commercially unviable for short-reach optical fiber links due to their high cost, footprint, and latency. Legacy systems employing intensity modulation and direct detection (IM/DD) with limited signal processing capabilities can provide low-cost solutions for inter-datacenter applications, however they cannot scale with the ever-increasing capacity requirements in modern communication networks. This has motivated renewed research interest in low-cost and low-complexity transceivers with bitrates optimized over short transmission distances (10-100 km) where channel impairments are dominated by dispersion with some nonlinear distortion \citep{106chagnon2019optical}.

\begin{figure}[h!]
\centering
\includegraphics[width=\linewidth]{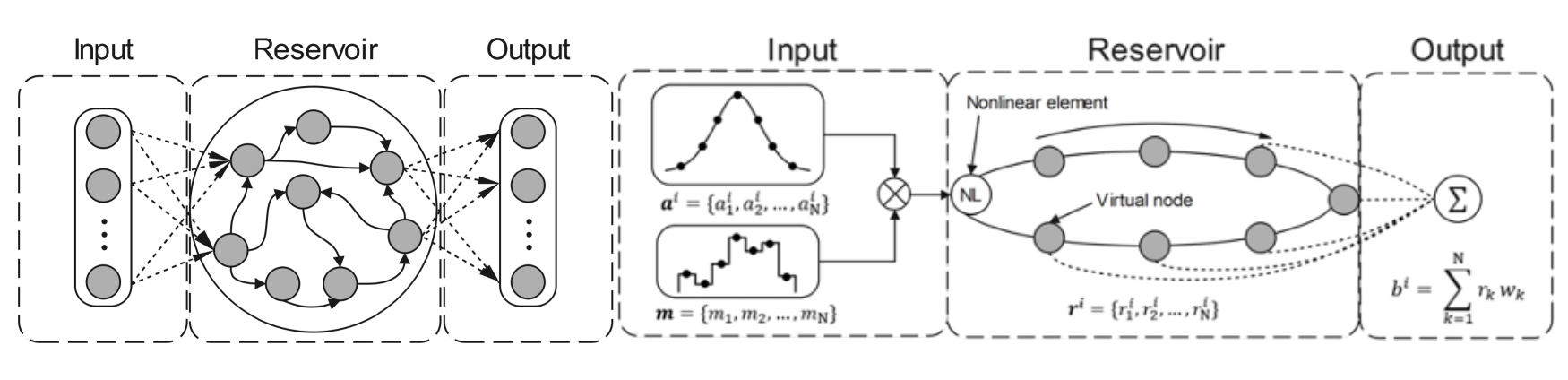}
% \vspace{-10pt}
\caption{(a) Generic model of a reservoir computer (b) Illustration of the RC equalization scheme in \cite{111argyris2018photonic} based on a single nonlinear node with time delayed feedback. The input $\bf{a}^i$ is a vector of N samples representing a single bit or symbol. The mask vector $\bf{m}$ , which defines the input weights, is multiplied by the elements in $\bf{a}^{i}$  and injected sequentially into the reservoir. The output is a linear combination of the virtual nodes in the reservoir with weights optimized to produce an estimate $\bf{b}^i$  of the bit or symbol value.}
\label{fig:figure3}
\end{figure}

Photonic neural networks based on reservoir computing techniques have shown promising results for channel equalization in fiber-optic links. Reservoir computers (RC) are a class of recurrent neural networks that consist of a reservoir of sparsely connected neurons with randomized fixed weights. Contrary to feed-forward recurrent neural networks which are trained using backpropagation or Hessian-free optimization, reservoir computers only require the output weights to be trained, which can be achieved by linear regression. Optical RCs have attracted significant research interest as the reservoir can be realized by a single nonlinear element with a delayed feedback loop \citep{107appeltant2011information} which has size- and cost-efficient implementations in photonic circuits. The first demonstrations of this technology for signal equalization tasks in FOCS used a semiconductor laser as a nonlinear element with a fiber delayed-feedback line and showed results competitive with DSP-based techniques \citep{103argyris2018photonic}. This approach is illustrated in Figure~\ref{fig:figure3}. Real-time operation of this photonic RC faces challenges, however, as the input layer is time-multiplexed by electronically masking each bit before injection into the reservoir, which incurs a speed penalty as the bit time must be stretched to match the delay of the feedback loop \citep{108sorokina2019fiber}. An all-optical implementation of a dual quadrature RC has also been proposed that could enable high bandwidth signal processing for coherent optical receivers \citep{108sorokina2019fiber}. In this design, nonlinear transformation of the input signal is achieved through the Kerr effect in a highly nonlinear fiber with a modulated pump to select the desired signal quadrature. Optoelectronic approaches have also been considered, with optical pre-processing via spectral slicing to improve the dynamics of a digital reservoir computer. This architecture addresses the system losses incurred in all-optical RCs and achieves significant reach extension compared to legacy IM/DD systems at the cost of higher complexity as the number of photodetectors scales linearly with the number of spectral slices \citep{109da2020reservoir}. Despite the SNR penalties, photonic RCs have merit over conventional DSP-based linear techniques when there is significant nonlinear distortion in the channel. Increasing the launch power at the transmitter may offset the system losses to achieve a higher OSNR with improved performance in the presence of nonlinear impairments compared to linear equalizers. This hypothesis is supported by \citep{110li2020photonic} which compared the BER vs OSNR trade-off for a state-of-the-art DSP based equalizer against a photonic RC in a 100 km 56 Gbd DWDM transmission system. As expected, the DSP equalizer is the better choice at low OSNR values, however the results show the RC-based equalization outperforms the digital receiver at high OSNR where the nonlinear perturbations are strong. 

% \subsection{Wireless communications}
\subsubsection{Jamming avoidance response}

\begin{figure}[h!]
\centering
\includegraphics[width=\linewidth]{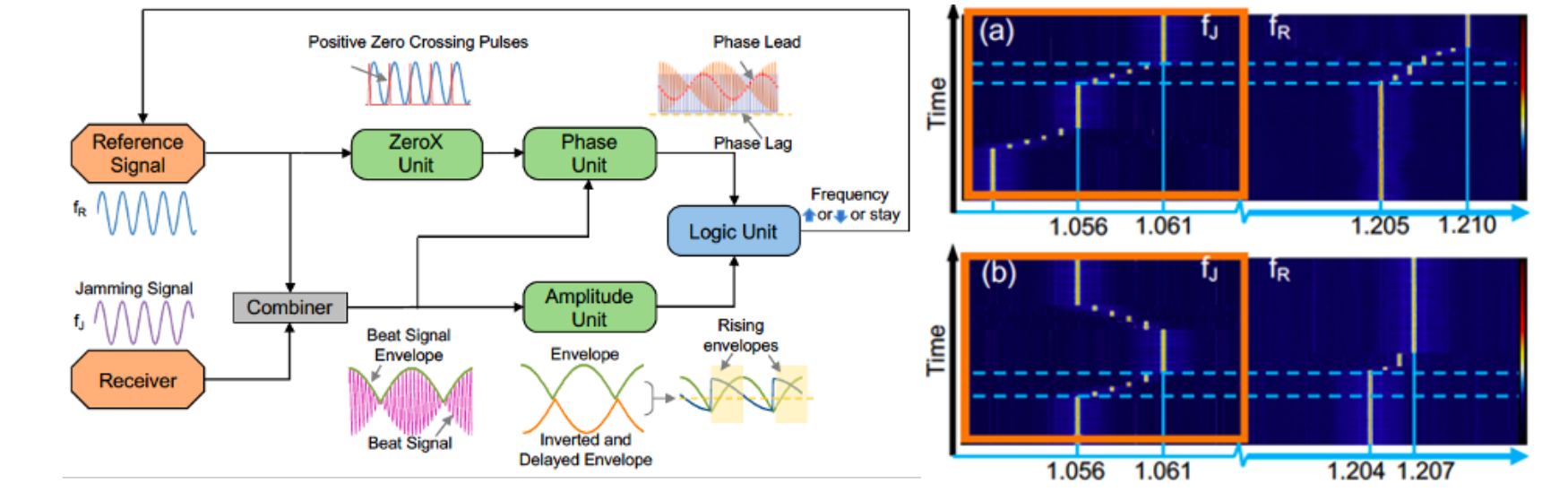}
% \vspace{-10pt}
\caption{(a) Illustration of the JAR design and the four functional units. (b) Spectral waterfall measurement of the photonic JAR in action with sinusoidal reference signal fR and jamming signals fJ = 150 MHz. (i) fJ is approaching fR from the low frequency side and triggers the JAR, (ii) fJ is approaching fR from the low frequency side and triggers the JAR, and then is moved away.}
\label{fig:figure4}
\end{figure}

The dramatic increased demand in mobile RF systems has significantly worsened the spectral scarcity issue, the overcrowded RF spectrum increases the likelihood of inadvertent jamming \citep{112wilhelm2011short,113poisel2011modern}. Inadvertent jamming is one type of jamming that comes from a friendly source, that is usually aimless and unforeseeable. However, inadvertent jamming could easily corrupt the transmission channel if not being mitigated properly. In fact, inadvertent jamming does not only happen in our communication systems. Eigenmannia, a genus of glass knifefishes uses electric discharge to communicate with their own species and to recognize different species. Since Eigenmannia does not have FCC to regulate their frequency allocation, their frequency usage is dynamic. To avoid jamming, the Eigenmannia has an effective Jamming Avoidance Response (JAR) that helps the fish to identify potential jamming and automatically move their electric discharge frequency away from the potential jamming frequency \citep{114bullock1972jamming,115scheich1977neural}. The JAR in Eigenmannia mainly consists of four functional blocks (Figure~\ref{fig:figure4}(a), they are (i) Zero-crossing point detection unit, (ii) Phase unit, (iii) Amplitude unit, and (iv) Logic unit. 

First, the ZeroX unit identifies the positive zero crossing points in the reference signal. Then, phase comparison between the reference signal and the beat signal takes place at the Phase unit. Amplitude unit takes the envelope of the beat signal and marks the rising and falling amplitudes differently. Lastly, the Logic unit takes the phase and amplitude information obtained and determines if the electric discharge frequency should be increased or decreased to avoid jamming, and if there is no potential jamming threat. 

This powerful JAR can be implemented using photonics to allow the JAR to be used in our communication frequency range \citep{116toole2016photonic,117fok2018photonic,118lin2018biomimetic}. The major device to achieve JAR is a semiconductor optical amplifier (SOA), where self-phase modulation is used in the Zero-crossing point detection unit, cross-gain modulation is used in both the Phase unit and amplitude units \citep{117fok2018photonic,118lin2018biomimetic}. As the jamming frequency is moving closer to the jamming range, the JAR will be activated and move the emitting frequency away gradually until it is out of the jamming range, as illustrated in the spectral waterfall measurement in Figure~\ref{fig:figure4}(b). The photonic JAR works well for frequencies from hundreds of MHz to tens of GHz, that provides an adaptive and intelligent solution to inadvertent jamming in emerging communication systems.

\subsubsection{Multivariate photonics - PCA, BSS, MIMO}

\begin{figure}[h!]
\centering
\includegraphics[width=0.6\linewidth]{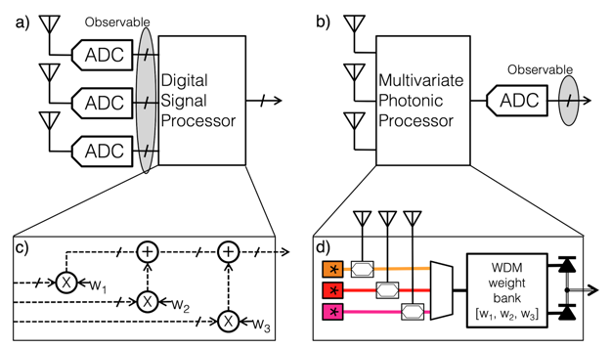}
% \vspace{-10pt}
\caption{(Comparison of multi-antenna radio front-ends followed by dimensionality reduction. a) Dimensionality reduction with electronic DSP in which each antenna requires an ADC. b) Dimensionality reduction in the analog domain (a.k.a physical layer) in which only one ADC is required. A photonic implementation of weighted addition is pictured, consisting of electrooptic modulation, WDM filtering, and photodetection. From \cite{tait2018blind}}
\label{fig:figure5}
\end{figure}

Multi-antenna systems provide an orthogonal dimension with which to share the electromagnetic spectrum. In many cases, interference can originate from sources that are either broad spectrum or frequency hopping. They cannot be eliminated by changing center frequency or filtering frequency. This interference can be separated by analyzing correlations in signals received by different antennas in an array. Photonic devices excel at processing radio-frequency (RF) signals. The radio signal is multiplexed onto an optical carrier that effectively serves as a much higher intermediate frequency, as shown in Figure~\ref{fig:figure5}. The resulting signal has a small fractional bandwidth, even for radio signals spanning 10s of GHz, which means that linear processing elements have low dispersion across the band. Once processed in the optical domain, the signal is converted back to RF by photodetection.

\subsubsection{Compressive sensing}
Compressive sensing is a paradigm for signal acquisition that exploits the inherent sparsity in many natural and man-made signals to enable sampling at rates far below the Nyquist frequency with limited or no loss of information. CS theory relies on linear dimensionality reduction to produce an efficient representation of an input signal by projecting it onto a rank-deficient sensing matrix. Provided that the input signal is sparse under some basis and the sensing matrix satisfies the requisite properties for sparse signal recovery, the input signal can be reconstructed by linear programming with polynomial complexity \citep{119duarte2011structured}. Compressive sensing has motivated the design of converters that can sense a wideband spectrum at sub-Nyquist sampling rates by performing the linear projections in the analog domain before digitization. The analog front end can be implemented by modulating the input signal with one or more periodic chip sequences followed by a low pass filter. This approach relaxes the bandwidth requirements of the ADC, however the RF front end must still operate at the Nyquist frequency to avoid aliasing. When the observation bandwidth is large, non-idealities in the chip sequences such as timing offset and insufficient slew rate can cause measurement errors similar to the clock jitter and aperture error that can occur in high speed ADCs \citep{120abari2013analog}. Several photonic implementations of a compressive sensing receiver have been proposed as a solution to the bandwidth limitations of analog electronics \citep{121wang2019photonic,122shmel2017photonic}. The high bandwidth offered by photonic pseudo random bit sequence generators and modulators enables analog domain compression of input signals with extremely wide bandwidths of 40 GHz or more that would be infeasible or impossible with electronic hardware. Furthermore, wavelength division multiplexing can be used to dynamically configure parallel sensing channels that share the same waveguide and modulator, which may offer cost and size benefits over electronic implementations\citep{123shaver2016photonic}.

\subsection{AI/Machine learning}

\subsubsection{Vector-matrix multipliers}

\begin{figure}[h!]
\centering
\includegraphics[width=0.6\linewidth]{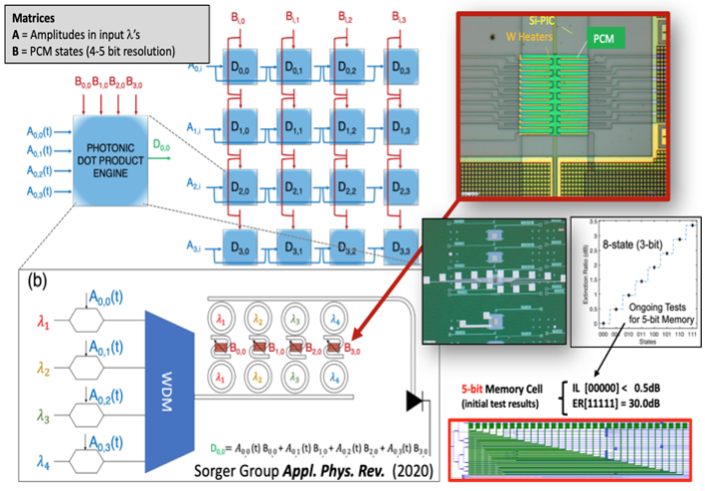}
% \vspace{-10pt}
\caption{Photonic Tensor Core (PTC) featuring POPS throughput and 10s ps-short latency to process VMM in PICs. (left) schematic of a modular exemplary PTC composed of 4$\times$4 photonic dot-product engines using RRs for WDM. Data entered via (exemplary) EO modulators at 50 Gbps are MUXed then dropped at passive RRs. Being 'passive' for the RR is importantly beneficial since it allows for reduced complexity, less real-estate used on the chip, and does not consume power unlike thermally-tuned approaches to perform the optical multiplication. The latter can be elegantly executed using nonvolatile programmable P-RAM elements using non-GST with a very low optical loss in the amorphous state \citep{124miscuglio2020artificial}}
\label{fig:figure6}
\end{figure}

With an ongoing trend in computing hardware towards increased heterogeneity, domain-specific coprocessors are emerging as alternatives to centralized paradigms. The tensor core unit has shown to outperform graphic process units by almost 3-orders of magnitude enabled by higher signal throughout and energy efficiency. In this context, photons bear several synergistic physical properties while PCMs allow for local nonvolatile mnemonic functionality in these emerging distributed non von-Neumann architectures. While several photonic neural network designs have been explored, a photonic tensor core to perform matrix vector multiplication and summation is yet to be implemented. An integrated photonics-based tensor core unit can be designed by strategically utilizing i) a photonic parallelism via wavelength division multiplexing, ii) high 2 Peta-operations-per second throughputs enabled by 10s of picosecond-short delays from optoelectronics and compact photonic integrated circuitry, and iii) near-zero power-consuming novel photonic multi-state memories based on PCMs featuring vanishing losses in the amorphous state. Combining these physical synergies of material, function, and system, we show, supported by numerical simulations, that the performance of this 4-bit photonic tensor core unit can be one order of magnitude higher for electrical data, whilst the full potential of this photonic tensor processor is delivered for optical data being processed, where we find a 2-3 orders higher performance (operations per joule) as compared to an electrical tensor core unit whilst featuring similar chip areas. This work shows that photonic specialized processors have the potential to augment electronic systems and may perform exceptionally well in network-edge devices in the looming 5G networks and beyond. (Figure.~\ref{fig:figure6})

\subsubsection{Convolutions (inference accelerator)}

% \todo{Brunner's introduction of convolutions suits here better, so I moved it up}

Convolutions of an image with a filter, i.e. the convolution kernel, are among the most heavily employed operations in computational imaging. During convolution, a kernel slides across an image and at each position the overlap integral between image and kernel provides the convolution’s value. This technique is extremely efficient in identifying entire objects \citep{125ambs2010optical} or only local features.  It therefore comes at not too much of a surprise that functional topologies implementing convolutions similar to Gabor filters have been identified in the visual cortex of mammals \citep{126olshausen1996emergence}. Convolutional neural networks (CNNs) leverage these concepts in a modern information processing context, and an image classification performance superior to humans makes CNNs relevant for a wide range of applications \citep{lecun2015deep}.

% \todo{I also slightly changed the wordings and added subheadings to make the transitions of these paragraphs more smooth}

\begin{figure}[h!]
\centering
\includegraphics[width=0.6\linewidth]{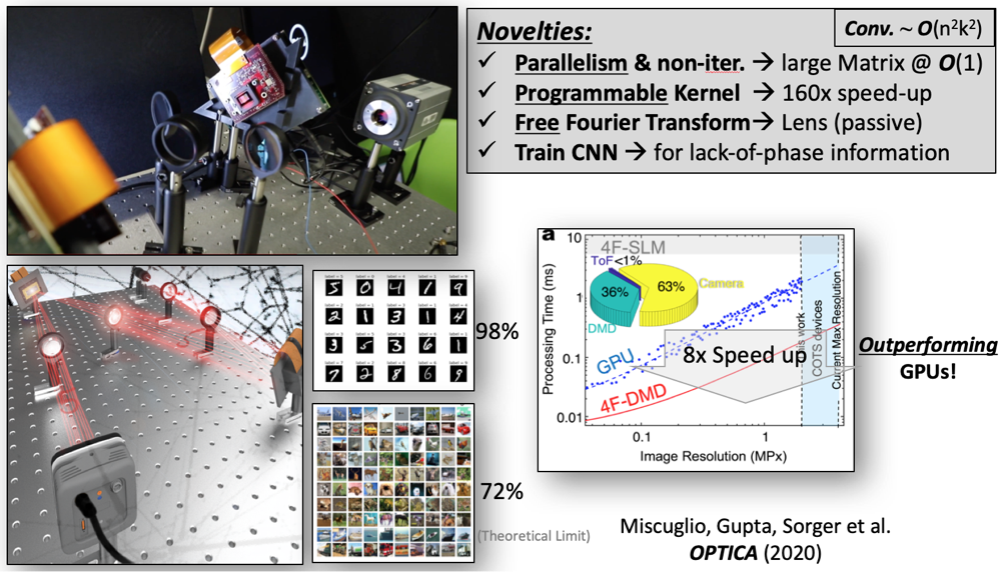}
% \vspace{-10pt}
\caption{Example of an Optical Convolutional Neural Network (CNN) accelerator exploiting the massive (10\textsuperscript{6} parallel channel) parallelism of free-space optics. The convolutional filtering is executed as point-wise dot-product multiplication in the Fourier domain. The conversion into and out-of the Fourier domain is performed elegantly and completely passively at zero power  While SLM-based systems can perform such Fourier filtering in the frequency domain, the slow update rates does not allow them to outperform GPUs. However, replacing SLMs with fast 10s kHz programmable digital micromirror display (DMD) units, gives such optician CNNs an edge over the top-performing GPUs. Interestingly, the lack-of-phase information of these amplitude-only DMD-based optical CNNs can be accounted for during the NN training process. For details refer to \cite{miscuglio2020massively}}
\label{fig:figure7}
\end{figure}

\textbf{Free-space optical convolution:} A fundamental aspect of computational imaging is that the primary information is distributed in two dimensions (2D). Free-space optical convolution setups inherently respect this encoding principle. Task-specific accelerators based on free-space optics bear fundamental homomorphism for massively parallel and real-time information processing given the wave-nature of light. However, initial results are frustrated by data handling challenges and slow optical programmability. Sorger's group recently introduced an amplitude-only Fourier-optical processor paradigm capable of processing large-scale $\sim$(1000 $\times$ 1000) matrices in a single time-step and 100 microsecond short latency \citep{miscuglio2020massively} (Figure~\ref{fig:figure7}). Conceptually, the information-flow direction is orthogonal to the two dimensional programmable-network, which leverages 10\textsuperscript{6} -parallel channels of display technology, and enables a prototype demonstration performing convolutions as pixel-wise multiplications in the Fourier domain reaching peta operations per second throughputs. The required real-to-Fourier domain transformations are performed passively by optical lenses at zero-static power. We exemplary realize a convolutional neural network (CNN) performing classification tasks on 2-Megapixel large matrices at 10 kHz rates, which latency outperforms current GPU and phase-based display technology by one and two orders of magnitude, respectively. Training this optical convolutional layer on image classification tasks and utilizing it in a hybrid optical-electronic CNN, shows classification accuracy of 98\% (MNIST) and 54\% (CIFAR-10). Interestingly, the amplitude-only CNN is inherently robust against coherence noise in contrast to phase-based paradigms and features an over 2 orders of magnitude lower delay than liquid crystal-based systems. Beyond contributing to novel accelerator technology, scientifically this amplitude-only massively-parallel optical compute-paradigm can be far-reaching as it de-validates the assumption that phase information outweighs amplitude in optical processors for machine-intelligence, such as for information processing at the network-edge, in data centers, or for pre-processing information or filtering towards near real-time decision making.

%\todo{Bhavin, do you want to write your work here}
\textbf{2D integrated optical convolution :}  Deep neural networks are based on CNNs which are powerful and highly ubiquitous tools for extracting features from large datasets for applications such as computer vision and natural language processing. The success of CNNs for large-scale image recognition has stimulated research in developing faster and more accurate algorithms for their use. However, CNNs are computationally intensive and therefore results in long processing latency. One of the primary bottlenecks is computing the matrix multiplication required for forward propagation. In fact, over 80\% of the total processing time is spent on the convolution~\citep{Li:2016}. Therefore, techniques that improve the efficiency of even forward-only propagation are in high demand and researched extensively \citep{Jaderberg:2014,Goodfellow:2016}. Recently, there has been much investigation of implementing convolution operations with integrated optics \citep{bangari2019digital, 82miscuglio2020photonic,Xu2021,Feldmann2021}. These approaches can speed up convolution operations by orders of magnitude (over current electronic processors such as GPUs and TPUs) by implementing fast and parallel vector-matrix multiplications with wavelength multiplexing techniques.

\begin{figure}[h!]
\centering
\includegraphics[width=0.6\linewidth]{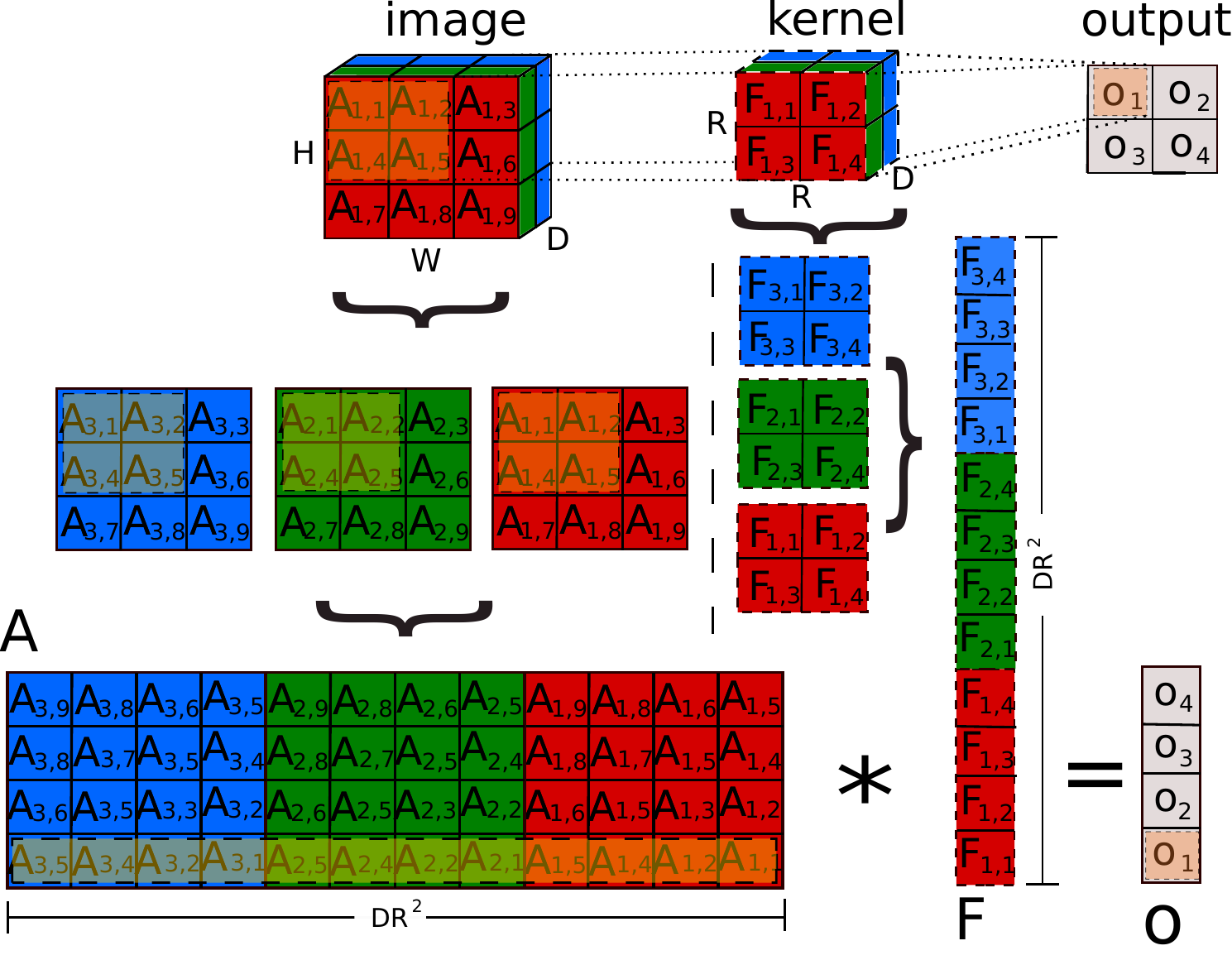}
% \vspace{-10pt}
\caption{Schematic illustration of a convolution. At the top of the figure, an input image is represented as a matrix of numbers with dimensionality $H\times W\times D$ where $H$, $W$ and $D$ are the height, width and depth of the image, respectively. Each element $A_{i,j}$ of $A$ represents the intensity of a pixel at that particular spatial location. The kernel $F$ is a matrix with dimensionality $R\times R\times D$, where each element $F_{i,j}$ is defined as a real number. The kernel is slid over the image by using a stride $S$ equal to one. As the image has multiple channels (or depth) $D$, the same kernel is applied to each channel. Assuming $H = W$, the overall output dimensionality is $(H - R + 1)^{2}$. The bottom of the figure shows how a convolution operation generalized into a single matrix-matrix multiplication.  where the kernel F is transformed into a vector $\mathbb{F}$ with $DR^{2}$ elements, and the image A is transformed into a matrix $\mathbb{A}$ of dimensionality $DR^{2} \times (H - R + 1)^{2}$. Therefore, the output is represented by a vector with $(H - R + 1)$ elements.}
\label{fig:convolutions}
\end{figure}

A convolution is a weighted summation of two discrete domain functions $f$ and $g$: $(f * g) = \sum_{t=-\infty}^{\infty}f[\tau] g[t-\tau]$ where $(f * g)$ represents a weighted average of the function $f[\tau]$ when it is weighted by $g[-\tau]$ shifted by $t$. The weighting function $g[-\tau]$ emphasizes different parts of the input function $f[\tau]$ as $t$ changes. Convolutions are well known to perform a highly efficient and parallel matrix multiplication using kernels ~\citep{Goodfellow:2016}. As depicted in Fig.~\ref{fig:convolutions} convolution of an image $\mathbb{A}$ with a kernel $\mathbb{F}$ that produces a convolved image $\mathbb{O}$. An image is represented as a matrix of numbers with dimensionality $H \times W \times D$, where $H$ and $W$ are the height and width of the image, respectively; and $D$ refers to the number of channels within the input image. Each element of a matrix $\mathbb{A}$ represents the intensity of a pixel at that particular spatial location. A kernel is a matrix $\mathbb{F}$ of real numbers with dimensionality $R \times R \times D$. The value of a particular convolved pixel is defined by: $O_{i,j} = \sum_{h=1}^{D}\sum_{q=0}^{iS+R}\sum_{p=0}^{jS+R}B_{q,p,h}A_{iS+q,jS+p,h}$, where $S$ is the stride” of the convolution. The dimensionality of the output feature is: $\lceil\frac{H-R}{S}+1\rceil \times \lceil\frac{W-R}{S}+1\rceil \times K$, where $K$ is the number of different kernels of dimensionality $R \times R \times D$ applied to an image, and $\lceil \cdot \rceil$ is the ceiling function. The efficiency of convolutions for image processing is based on the fact that they lower the dimensions of the outputted convolved features. Since kernels are typically smaller than the input images, the feature extraction operation allows efficient edge detection, therefore reducing the amount of memory required to store those features.

In 2014, Tait et al.~\citep{tait2014broadcast} a proposed a scalable silicon photonic neural network called "broadcast-and-weight" (B\&W) which was demonstrated in 2017~\citep{tait2017neuromorphic} concurrently with other silicon photonic neuromorphic architectures~\citep{shen2017deep,shainline2017superconducting}.  Since the B\&W architecture is based on wavelength mutiplexing, parallel matrix multiplication can be leveraged to perform operations such as convolutions. In B\&W architecture, WDM signals are weighted in parallel by a bank of MRRs used as tunable filters~\cite{Tait18,huang2020demonstration}, summed with balanced photodiodes (BPD), and nonlinear activation functions implemented with MRR modulators. In 2020, based on this architecture, Bangari et al.~\citep{bangari2019digital} introduced a digital electronic and analog photonic (DEAP) architecture capable of performing highly efficient CNNs.

\begin{figure}[h!]
\centering
\includegraphics[width=1.0\linewidth]{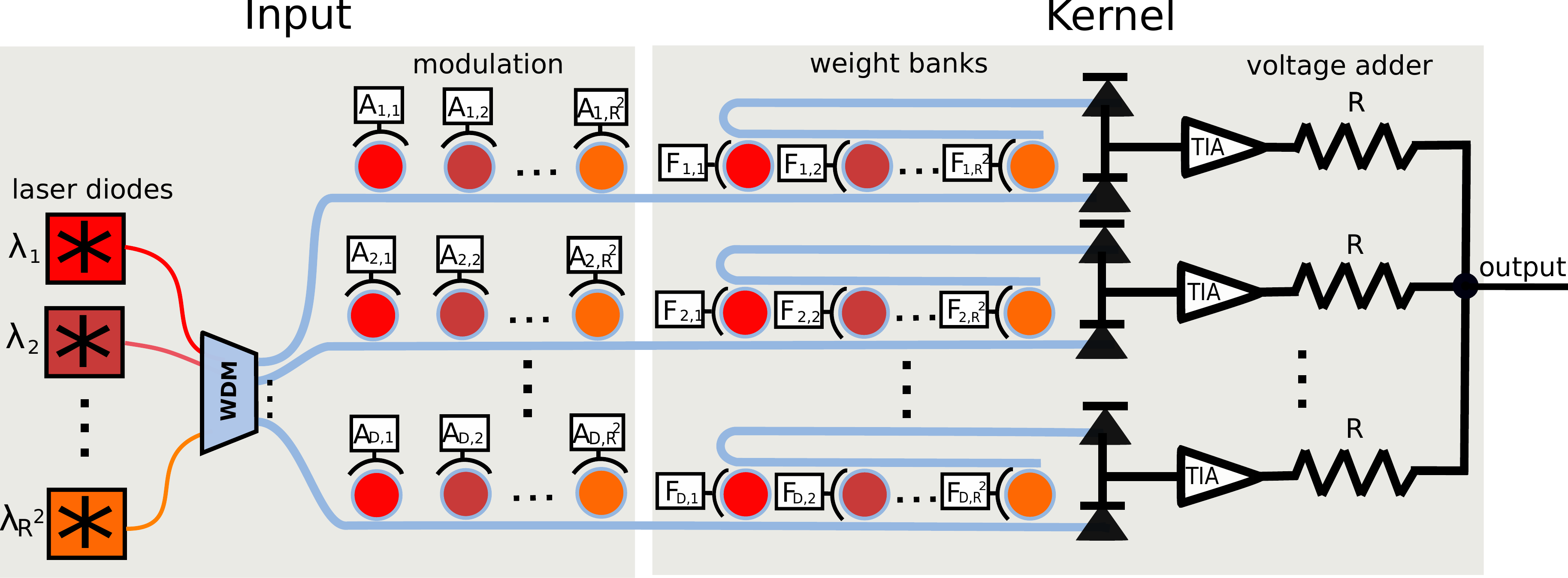}
% \vspace{-10pt}
\caption{Photonic architecture for producing a single convolved pixel. Input images are encoded in intensities $A_{l,h}$, where the pixel inputs $A_{m,n,k}$ with $m\in[i,i+R_{m}],n\in[j,j+R_{m}],k\in[1, D_m]$ are represented as $A_{l,h}$, $l = 1,\ldots, D$ and $h = 1,\ldots, R^{2}$. Considering the boundary parameters, we set $D = D_{m}$ and $R = R_{m}$. Likewise, the filter values $F_{m,n,k}$ are represented as are represented as $F_{l,h}$ under the same conditions. We use an array of $R^{2}$ lasers with different wavelengths $\lambda_{h}$ to feed the MRRs. The input and kernel values, $A_{l,h}$ and $F_{l,h}$ modulate the MRRs via electrical currents proportional to those values. Then, the photonic weight banks will perform the dot products on these modulated signals in parallel. Finally, the voltage adder with resistance $\mathbf{R}$ adds all signals from the weight banks, resulting in the convolved feature.}
\label{fig:deap_architecture}
\end{figure}

Fig.~\ref{fig:deap_architecture} shows the silicon photonic implementation of DEAP for performing convolution operations. To handle convolutions for kernels with dimensionality up to $R \times R \times D$, $R^2$ lasers are required with unique wavelengths since a particular convolved pixel can be represented as the dot product of two $1 \times R^2$ vectors. To represent the values of each pixel, $DR^2$ add-drop modulators (one per kernel value) are required where each modulator keeps the intensity of the corresponding carrier wave proportional to the normalized input pixel value. The $R^2$ lasers are multiplexed together using WDM, which is then split into $D$ separate lines. On every line, there are $R^2$ add-drop MRRs (where only input and through ports are being used), resulting in $DR^2$ MRRs in total. Each WDM line will modulate the signals corresponding to a subset of $R^2$ pixels on channel $k$, meaning that the modulated wavelengths on a particular line corresponds to the pixel inputs $(A_{m,n,k})$; $m\in[i,i+R], n\in[j,j+R]$ where $k \in [1, D]$, and $S = 1$. The $D$ WDM lines are then be fed into an array of $D$ photonic weight banks (WB). Each WB contains $R^2$ MRRs with the weights corresponding to the kernel values at a particular channel. Each MRR within a WB is tuned to a unique wavelength within the multiplexed signal. The outputs of the weight bank array are electrical signals, each proportional to the dot product $(F_{m,n,k}) \cdot (A_{p,q,k})$; $m\in[1,R^2 ]$, $n\in[1,R^2 ]$, $p\in[i,i+R^2 ]$, $q\in[j,j+R^2 ]$, where $k \in[1, D]$, and $S = 1$. Finally, the signals from the weight banks are added together. This can be achieved using a passive voltage adder. The output from this adder will therefore be the value of a single convolved pixel.

Recently, Feldman et al.~\citep{Feldmann2021} experimentally demonstrated this approach with a photonic tensor core for parallel convolutional processing achieving bandwidths of  2~TMACs/s and compute densities of 555~GMACs/s/mm$^2$with 5-bits of precision. Their system consists of a $3\times 3$ filter, 4~channels and filters and 14~GHz modulation and detection bandwidth, with a MAC cell area of 285 $\mu$m $\times$354 $\mu$m. With improved devices, efficiency, bandwidth, and integration densities, such a tensor core could feature a computational bandwidth of 1~PMACs/s and compute density of 15.6~TMACs/s/mm$^2$ with 50~GHz modulation and detection bandwidths, tensor core size of 50$\times$50, and 187 wavelength channels, and MAC cell area of 30 $\mu$m $\times$ 30 $\mu$m. For comparison, the Google TPU~\citep{10.1145/3079856.3080246} has a compute density of 150 GMACs/s/mm$^2$ with 8-bits of precision.

\begin{figure}[h!]
\centering
\includegraphics[width=1\linewidth]{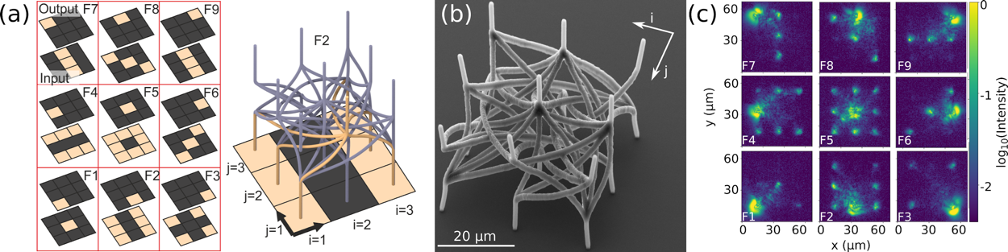}
% \vspace{-10pt}
\caption{Fully parallel and passive convolutions integrated in 3D. (a) 9 convolutional Haar filters and their 3D integration topology using photonic waveguides. (b) SEM micrograph of the 3D printed convolutional filter unit, fabricated with direct laser writing. (c) The filter’s optical transfer function agrees well with the target defined in (a). Figure reproduced from \cite{128moughames2020three}.}
\label{fig:figure8}
\end{figure}

\textbf{3D integrated optical convolution:} Information is kept within 2D spatial encoding, while light propagation through passive components along a third dimension implements the required operation, i.e. Fourier and inverse Fourier transforms as well as multiplication with a filter kernel. Novel 3D nano-fabrication \citep{127deubel2004direct} can now create intricate photonic waveguide circuits \citep{128moughames2020three} and holograms \citep{129dinc2020computer} that equally leverage the primary encoding space of images. Figure~\ref{fig:figure8}(a) shows how Boolean convolutional topologies can be 'hard-wired' in 3D. This intricate 3D routing can then be realized using 3D photonic waveguides, fabricated using two-photon polymerization of femto-second laser pulses \citep{127deubel2004direct}, and Figure~\ref{fig:figure8}(b) shows a SEM micrograph, and the resulting convolution filter’s transfer function, Figure~\ref{fig:figure8}(c), agrees well with the target. Importantly, 3D integration makes such interconnects scalable in size \citep{130dinc2020optical}, and motivated by identical scalability arguments similar 3D integration is already explored in electronics \citep{lin2020three}. However, such electronics chips will face severe thermal management challenges due to capacitive energy deposition into a volumetric circuit, and photonics therefore has an inherent advantage towards scalable integration of parallel convolutional filters.

\subsubsection{Edge or fog computing: image processing and super resolution filtering}

Photonic neural networks (e.g. \cite{82miscuglio2020photonic}) and free-space Fourier-optics 4f processors \citep{miscuglio2020massively} can be particularly beneficial to tasks that involve, for instance, Super-Resolution on Object Detection Performance in Satellite Imagery. Some modern camera sensors, present in everyday electronic devices like digital cameras, phones, and tablets, are able to produce reasonably high-resolution (HR) images and videos. The resolution in the images and videos produced by these devices is in many cases acceptable for general use. However, there are situations where the image or video is considered low resolution (LR). Examples include the following situations: (i) Device resolution limitation (as in some surveillance systems and satellite images). (ii) Object relatively small in a larger context; e.g. faces or vehicle located far away from the camera. (iii) Blurred or noisy images. being the device mounted on moving automated device (e.g. drones). (iv) Improving the resolution as a pre-processing step improves the performance of other algorithms that use the images; pattern recognition and target tracking. Super-resolution is a technique to obtain a high resolution (HR) image from one or several LR images. In this case one can use a training set to train (offline) a convolutional neural network (CNN) to learn to map between a low-resolution image and the high-resolution image within the training set; for instance, using a 4f-based system one can obtain pixel-wise dot product in the Fourier domain between the large input matrix (2MP$\times$) with a reprogrammable kernel that has been pre-patterned/written according to the training. The re-programmability, in a first instance, can be achieved using micromirror devices, which by shining light on top of the film, can locally change the phase of the PCM generating a 2D pattern. Although, this solution is limited by the speed at which the film can be written, therefore, in future implementations one can independently and simultaneously tune each pixel by changing their phases electrothermally in $\mu$s-time scale (requires additional circuitry), thus acting as 2D space filter in the Fourier domain or as layer of a convolutional network. Providing a forward looking view, if one would substitute the liquid crystal from the SLMs and the micromirrors from the DMDs with GHz-fast updating elements, but keep the same 1000 x 1000 pixel real-estate, the system would yet improve by a factor of $10^{10}$ to $10^8$
from SLM's and DMD's, respectively. Such electro-optic modulation, however, must be ultra-compact and regular photonics modulators based on carrier injection or depletion using Silicon or using the Pockel's effect in Lithium Niobate \cite{wang2018integrated} are non-usable due to millimeter to centimeter-large modulator device footprints. An alternative approach is to utilize higher index-changing emerging EO materials and heterogeneously integrate them with photonic waveguides. ITO and its ability to produce epsilon-near-zero \cite{amin2017active}, a nonlinearity enhancements (including EO nonlinearity) can be used to demonstrate micrometer-compact modulators, for example.  \cite{sorger2012ultra}, \cite{amin20180}, \cite{amin2020lateral}, \cite{amin2020sub}.

\subsubsection{Ultrafast learning and STDP}

What makes neuron fascinating is its ability to learn and adapt, which is a powerful capability that governs our actions, thoughts, and memories. These important functions of humans are relying on the synaptic plasticity between neurons, which is self-adjusted based on the information being processed and the response of the neuron itself. Among different synaptic weight plasticity models, spike timing dependent plasticity (STDP) is the most popular one, which is a biological process that adjusts the interconnection strength between neurons based on the temporal relationship (i.e. timing and sequence) between pre-synaptic and post-synaptic activities \citep{131song2000competitive}. The more you are running a certain neural circuit, the stronger the circuit becomes. 

\begin{figure}[h!]
\centering
\includegraphics[width=1\linewidth]{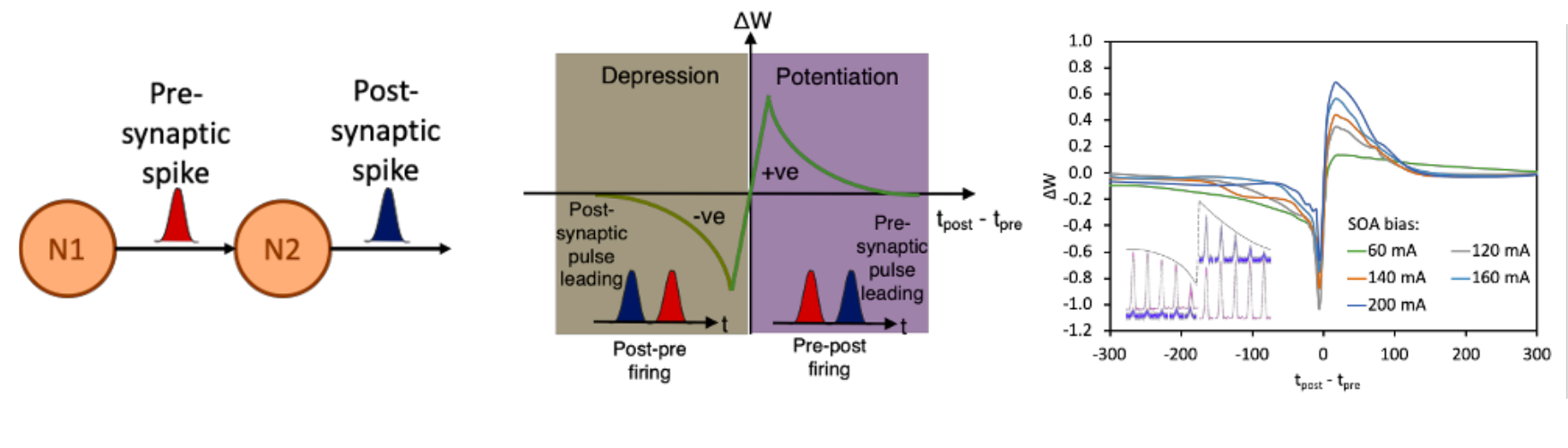}
% \vspace{-10pt}
\caption{(a) Illustration of pre-synaptic and post-synaptic spikes from two neurons N1 and N2. (b) Illustration of a STDP response. Right purple region: long-term potentiation; Left brown region: long-term depression. tpost-tpre: time difference between the firing of the post- and pre-synaptic spikes. (c) Experimentally measured STDP curves from the photonic based STDP circuit.}
\label{fig:figure9}
\end{figure}

In STDP, the interconnection strength between two neurons (N1 and N2) is determined by the relative timing and sequence between the presynaptic (red) and post-synaptic spikes (blue), as illustrated in Figure~\ref{fig:figure9}(a). If N2 spikes shortly after the stimulation from the presynaptic spike, the interconnection strength will be significantly increased and  results in long-term potentiation (LTP) of the connection strength, as illustrated by the shaded purple region in Figure~\ref{fig:figure9}(b). However, if N2 spikes before the stimulation from the pre-synaptic spike, the interconnection strength will be significantly decreased, resulting in long-term depression (LTD) of the synaptic connection, as illustrated by the shaded brown region in Figure~\ref{fig:figure9}(b). The exact amount of synaptic connection strength increment/decrement depending on the precise timing difference between the pre-synaptic and post-synaptic spikes of N2. 

To enable ultrafast learning in photonic neuron network, STDP has to be implemented using photonics and integrated into the photonic neuron network \citep{132fok2013pulse,133toole2015photonic}. One promising solution to implement STDP in photonic neuron network is using semiconductor optical amplifier (SOA) \citep{132fok2013pulse,133toole2015photonic,134toole2015photonic}. SOA has a unique gain dynamic that is sensitive to the timing and sequence of the input stimulations, which is similar to the STDP in neurons. One example is to use both cross-gain modulation and cross-polarization modulation in SOA to mimic the LTP and LTD responses in biological neurons \citep{134toole2015photonic}, the resultant photonic STDP is shown in Figure~\ref{fig:figure9}(c) that reassemble the biological STDP. It has been shown that supervised learning can be achieved using a SOA based STDP and two SOAs based neurons \citep{132fok2013pulse,133toole2015photonic}.

\subsubsection{Boolean learning via coordinate descent}

Current neural network optimization is largely based on back-propagating error gradients, and today’s boom in neural network applications is closely linked to the successful implementation of this concept in digital hardware. However, in hardware with unidirectional, i.e. forward flow of information, its implementation requires calculating the error-gradient for each network connection according to the chain rule of differentiation. In digital hardware this creates an enormous overhead, while in analog networks each weight and neuron parameter needs to be probed and stored, which ultimately is prohibitively complex in most settings. In systems where information can symmetrically propagate forward as well as backward, such as often the case in photonics, a regulatory error signal could be sent backwards. However, in order to physically implement weight optimization according to error back propagation using hardware rules such as STDP, a neuron’s nonlinearity in backward direction needs to be the gradient of its nonlinearity in forward direction. Such asymmetric neurons are a challenge that remains largely out of reach until the current day, and the only viable photonic concepts rely on phase conjugation \citep{135psaltis1987multilayer,136zhou2020situ}.

Coordinate descent is a practical as well as efficient alternative to error back propagation and currently is heavily explored in the field of machine learning. Individual or sets of weights i.e. coordinates, usually drawn at random, are modified in order to probe the error landscape's local gradient. Probing is followed by updating weights opposite to the gradient's direction and with a weighting factor dubbed the learning rate. In photonic hardware Boolean neural network connection weights have been realized via a digital micro-mirror device (DMDs) \citep{137bueno2018reinforcement}, and in a Boolean context the error gradient is probed by inverting a set of weights. Should the associated error gradient be negative, then the last modification is kept, otherwise it is discarded and the connections revert back to the previous configuration.

Boolean weights are currently explored in the general context of neural networks \citep{138courbariaux2016binarized} as well as in special-purpose electronic hardware \citep{139hirtzlin2020digital}. In photonic reservoir computing \citep{137bueno2018reinforcement} it was shown that Boolean coordinate descent converges exponentially and achieves chaotic signal prediction accuracy only slightly below a comparable photonic reservoir \citep{140bueno2017conditions} where double-precision weights were optimized offline. Furthermore, instead of a fully random, i.e. Makovian selection of descent coordinates, leveraging a greedy selection strategy allowed the system to converge twice as fast.

Boolean photonic weights implemented via a DMD enable programmable photonic neural networks comprising thousands of connections. In digitally emulated \citep{138courbariaux2016binarized}, electronically \citep{139hirtzlin2020digital} and photonically implemented \citep{137bueno2018reinforcement} neural networks, such binarized weights resulted only in slight performance penalties. DMD based Boolean coordinate descent in photonic neural networks therefore harbors great prospects for future, practical yet high performance neural networks, which noteworthy can be readily programmed based on classical software tools.

\subsection{Nonlinear programming}

\subsubsection{Solving optimization problems (model predictive control)}

Solving mathematical optimization problems lies at the heart of various applications present in modern technology such as machine learning, resource optimization in wireless networks, and drug discovery. Many optimization problems can be written as a quadratic program. For example, the least squares regression method can be mathematically mapped to a relatively easy quadratic program (with a positive definite quadratic matrix). Quadratic Programming refers to algorithms related to solving the optimization problem of finding the extremes of a quadratic objective function subject to linear constraints.  The general formulation of a quadratic program is usually solved iteratively, often requiring many time steps to reach the desired solution. The difficulty of quadratic programming grows exponentially with the dimension of the problem. Algorithms that can deal with large dimensions involve more computationally intensive techniques such as genetic algorithms or particle swarm optimization. As a result, conventional digital computers must either be limited to solving quadratic programs of very few variables, or to applications where the computation time is non-critical. This is why traditional computers are not appropriate to implement algorithms depending on QP for high-speed applications such as signal processing and control systems. In machine learning, many algorithms, such as SVM, require offline training because of the computational complexity of QP, but would be much more effective were they trained online.

A number of high-speed control problems, e.g. controlling plasma in aircraft actuations, fusion power plants, guiding of drones etc., are currently bottlenecked by the speed and latency of the control algorithms. Model predictive control (MPC) is an advanced technique to control complex systems, outperforming traditional PID control methods because it is able to predict control violations, rather than react to it. However, its control loop involves solving a quadratic problem at every control step, and therefore it is not computationally tractable for systems requiring speeds higher than kHz. Photonic neural networks help overcome this tradeoff by using techniques such as wavelength division multiplexing, which enables hundreds of high bandwidth signals (20 GHz) to be guided through a single optical waveguide.

Neural networks were demonstrated to solve general-purpose quadratic optimization problems by Hopfield and Tank \citep{141cichocki1993neural,142hopfield1986computing}. Until now, Hopfield networks have not been commonly implemented in hardware due to its all-to-all connectivity, which creates an undesirable tradeoff between neuron speed and neural network size --- in an electronic circuit, as the number of connections increases, the bandwidth at which the system can operate decreases \cite{tait2014broadcast}. This means a photonic Hopfield network implementation can simultaneously tackle quadratic programs with large dimensions and converge in nanoseconds \citep{de2019machine}. Implemented in photonic hardware, model predictive control can be employed in systems operating in the MHz regime.

\begin{figure}[h!]
\centering
\includegraphics[width=0.8\linewidth]{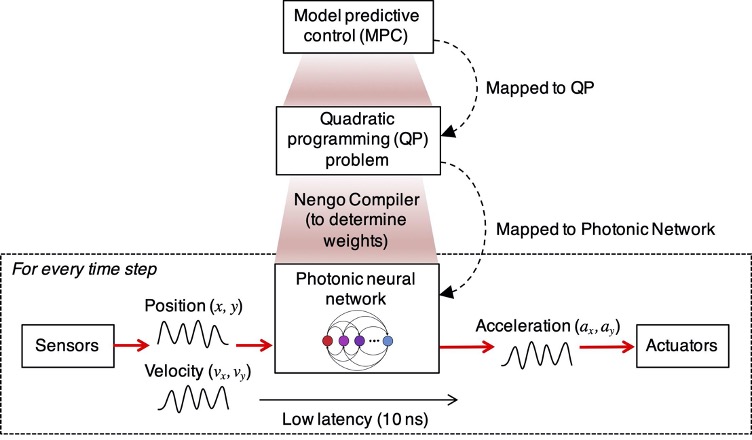}
% \vspace{-10pt}
\caption{Schematic figure of the procedure to implement the MPC algorithm on a neuromorphic photonic processor. Firstly, map the MPC problem to QP. Then, construct a QP solver with continuous-time recurrent neural networks (CT-RNN) \citep{141cichocki1993neural}. Finally, build a neuromorphic photonic processor to implement the CT-RNN. The details of how to map MPC to QP, and how to construct a QP solver with CT-RNN are given in \cite{de2019machine}. Adapted from \cite{de2019machine}.}
\label{fig:figure10}
\end{figure}

\subsubsection{Ordinary differential equation solving with a neural compiler}

While neural networks are often used for their learning properties, they can also be programmed directly. Direct programming of analog systems has a major pitfall in that the components are unreliable and subject to parameter variation. One approach is to represent variables as population states that are robust to parameter variations. The neural engineering framework (NEF) \citep{143stewart2014large} provides an algorithm to program ensembles of imperfect analog devices to perform operations on population coded variables.

\cite{tait2017neuromorphic} demonstrated a programmable network of two photonic neurons. The NEF algorithm was fed the responses of photonic devices, resulting in a weight matrix that allows the network to approximate variables, operations, and differential equations. It was shown in simulation how 24 neurons could emulate the Lorenz attractor. The approximation improves with more neurons.

The example of the Lorenz attractor is an example of a task-based benchmark for a 24-neuron network. Task-based benchmarks play an important role in validating experimental systems as they begin to incorporate more neurons and parameters. The example also demonstrated the compatibility between photonic neural networks and the NEF, including all of the NEF's key principles and consequent functionality. Using this neural compiler provides a route to a variety of known applications and other benchmarks \citep{144stewart2015closed}.

\begin{figure}[h!]
\centering
\includegraphics[width=0.8\linewidth]{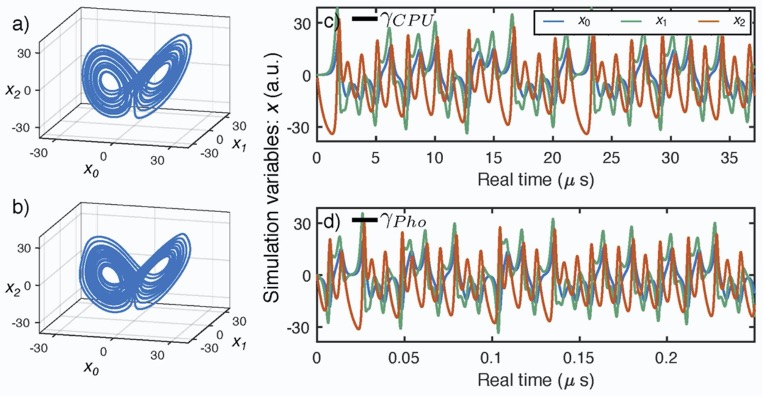}
% \vspace{-10pt}
\caption{Photonic neural network benchmarking against a CPU. (a,b) Phase diagrams of the Lorenz attractor simulated by a conventional CPU (a) and a photonic neural network (b). (c,d) Time traces of simulation variables for a conventional CPU (c) and a photonic CTRNN (d). The horizontal axes are labeled in physical real time, and cover equal intervals of virtual simulation time, as benchmarked by $\gamma$CPU and $\gamma$Pho. The ratio of real-time values of $\gamma$'s indicates a 294-fold acceleration. From \cite{tait2017neuromorphic}}
\label{fig:figure11}
\end{figure}

\subsection{Cryptography and security }
\subsubsection{Optical steganography (bio-inspired by Marine Hatchetfish camouflage strategies)}

Communication systems have been integrated deeply in our daily lives, supporting applications including online banking, augmented reality experience, and telemedicine. Therefore, it is expected that the massive amount of sensitive and personal information are needed to be protected against attacks. To provide information security, sophisticated encryption schemes are used at the higher layer of the communication system, i.e. media access control (MAC) layer. However, a physical layer without property security measures makes a communication system vulnerable to attack, resulting in total exposure of sensitive information \citep{146skorin2016physical}.

Effective physical cryptography requires both encryption and steganography. Encryption scrambles the sensitive information so that it is unreadable without the key, while steganography hides the sensitive information in plain sight so that the attacker will not even know there is sensitive information to look for. It is like hiding valuables in a locked safe (encryption) behind a secret bookcase door (steganography). Physical encryption schemes have been intensively studied, but physical layer steganography is still underdeveloped \citep{fok2011optical,149wu2013optical,148wang2010optical}. 

\begin{figure}[h!]
\centering
\includegraphics[width=1\linewidth]{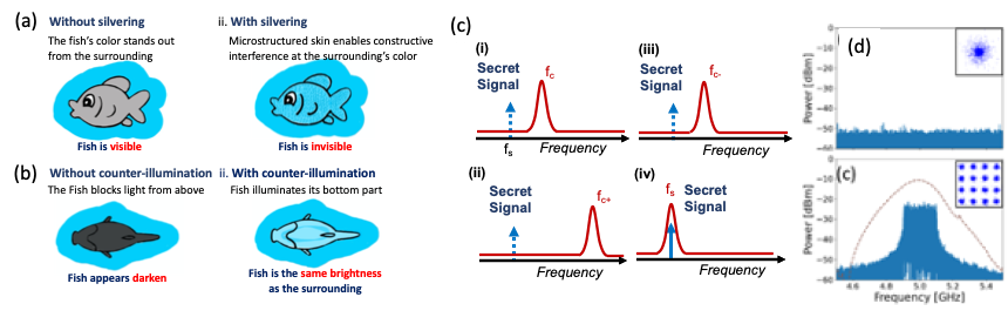}
% \vspace{-10pt}
\caption{a) Side view (i) no camouflage - fish is visible (ii) silvering - fish is destructively interfered at colors that could indicate the presence of the fish; (b) Bottom view (i) no camouflage – fish appears darker against the bright water surface when seen from below (ii) counterillumination – fish illuminates itself to the same color and intensity as the background. (c) Silvering (i) photonic RF FIR creates destructive interference condition at the stealth signal frequency (f\textsubscript{s}); (ii) Transmission in optical fiber will only push the constructive interference condition to a much higher frequency (f\textsubscript{c+}); (iii) Dispersion compensation fiber at the last section of the transmission will move the constructive interference condition back to f\textsubscript{c}; (iv) Correct dispersion at the stealth receiver allows constructive interference condition to occur at the stealth signal frequency f\textsubscript{s}. (d) Measured RF spectra and constellation diagrams (i) during transmission without a correct stealth receiver (ii) at the designated stealth receiver with correct location and dispersion. }
\label{fig:figure12}
\end{figure}

Turning to nature for an effective solution, Marine Hatchetfish is a master of hiding its appearance in the deep ocean using unique ocean camouflage techniques \citep{150rosenthal2017three}. Firstly, Marine Hatcherfish has microstructured skin on the sides so that only light that is similar to its surrounding is constructively interfered, while colors that could disclose its presence are destructively interfered, this technique is called silvering (Figure~\ref{fig:figure12}(a)). Secondly, Marine Hatchetfish also generates and directs light to the bottom part of its body to illuminate itself so that its color and brightness is the same as its surrounding when seen from below, this technique is called counter-illumination (Figure~\ref{fig:figure12}(b)). The two camouflage strategies allow the Marine Hatchetfish to conceal its appearance in all directions.

Borrowing the camouflage strategies from Marine Hatchetfish and applying it in RF signal transmission in optical fiber would be a natural and an effective way to achieve steganography. Silvering can be achieved using photonic FIR to make the sensitive signal disappear in the eyes of the attacker through destructive interference \citep{151liu2021bio}. Figure~\ref{fig:figure12}(c) shows the transformation of the FIR response at (i) the stealth transmitter, (ii) during and after signal transmission in single mode fiber, (iii) after dispersion compensation, and (iv) at the designated stealth receiver. The stealth signal is invisible at any point during the transmission and can only be retrieval with a precise stealth receiver at the designated location, Furthermore, counter-illumination can be achieved using a noise-like optical carrier that has the same spectral content and intensity as the system noise, similar to how Marine Hatchetfish illuminate itself \citep{151liu2021bio}. Steganography using bio-inspired silvering and counter-illumination techniques allow the sensitive signal to be concealed in all possible domains that the attacker could be looking at. Figure~\ref{fig:figure12}(d)i and ii shows the measured RF spectra and constellation diagrams of the stealth signal during transmission and at the designated stealth receiver. It is proven that the stealth signal is disappeared in the eyes of the attacker.

\subsubsection{Encryption and decryption}

\begin{figure}[h!]
\centering
\includegraphics[width=1\linewidth]{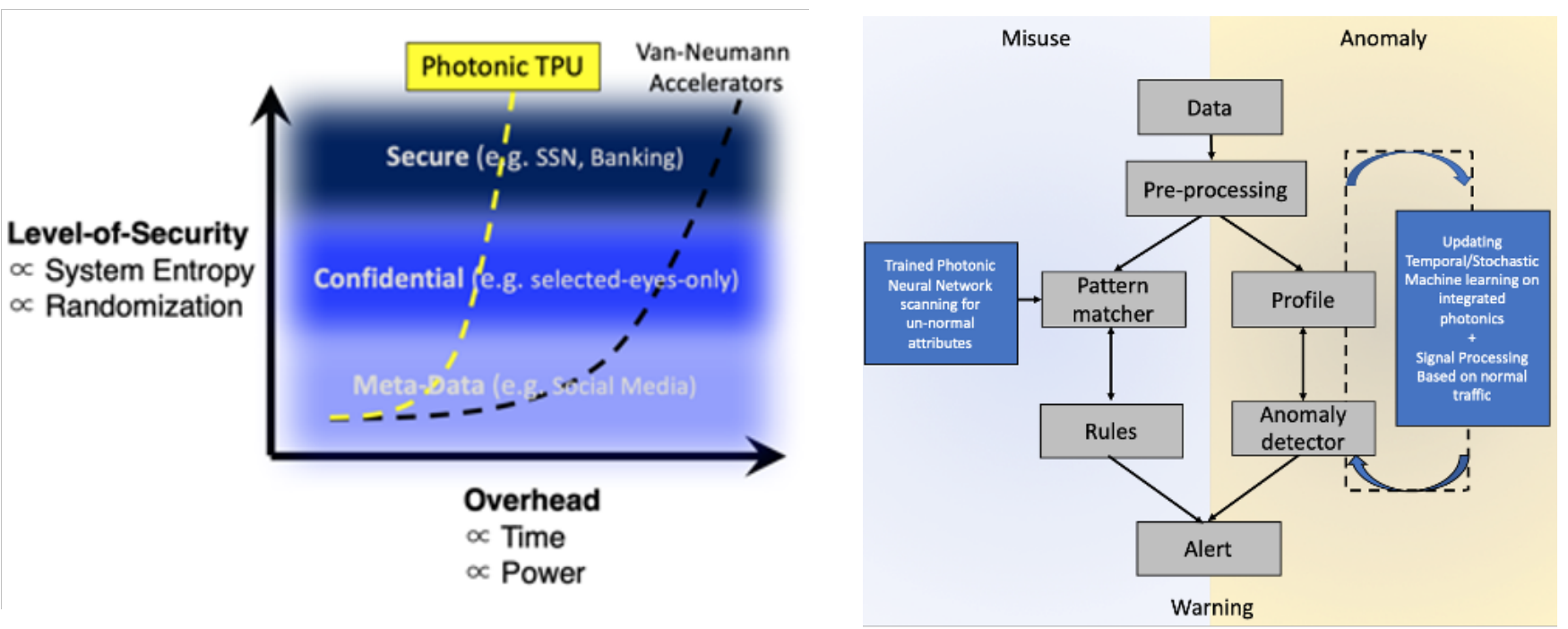}
% \vspace{-10pt}
\caption{(left) Photonic machine-learning accelerators such as photonic tensor core (PTU) processors \citep{82miscuglio2020photonic} enabled a higher level of data security with reduced overhead (e.g. time and power consumption). This opens possibilities for not only securing 'secure' and 'confidential' data, but also meta-data. (right) Flow-chart for detection of anomalies and misuses exploiting rapidly updating photonic neural network and trained photonic neural network which embeds on chip photonic memory, respectively. }
\label{fig:figure13}
\end{figure}

Movement of massive data traffic is rapidly growing in this big-data era, i.e. the 4\textsuperscript{th} industrial revolution of 'digitalization', spurred by the emergence of machine intelligence. Since the compute capacity at any one physical location is limited by the bounds of power and cooling requirements, it is naturally inevitable that data movement will be necessary. At the same time, it is critical to maintain information security during data transit between distributed computing locations. Encryption-in-transit mechanisms protect the integrity and confidentiality of sensitive information transmitted over Internet infrastructure between physical computing locations. However, due to the scale of data volumes ($\sim$Terabits/sec), efficient (en/de)cryption mechanisms are paramount for effectiveness of encryption-in-transit --- exactly what optical communication links provide. Since the data is already in the optical domain, thence, it is natural to consider (pre)processing information in the same optical and analog-domain (Figure \ref{fig:figure13}). That is, avoiding domain crossings such as optical-to-electrical (OE, vise-versa) and eliminating digital-to-analog (and vise-versa) not only saves energy-per-compute, but also improves system delay and reduces system complexity. The latter, is important for scaling vectors such as volume, reliability and ultimate cost of the system and hence the application. 

Data security applications include three main areas: \textbf{Authentication}, which is the verification of data sources and destinations. Such functions are carried out by modules embedded within Trusted Computing Platforms, which ensure that only legitimate users can send and receive data. Modern advances in multi-factor authentication mechanisms \citep{bhargav2007privacy} have incorporated several aspects such as knowledge factors, ownership factors and inherence factors. \textbf{Integrity}, which considers ensuring that the transmitted data arrives at the destination in an unmodified manner. Data integrity is usually guaranteed with checksum mechanisms that include error correcting codes. Such integrity-preserving mechanisms have been incorporated even inside modern hardware- for instance, some CPU architectures employ transparent checks to detect and mitigate data corruption in CPU caches, buffers and instruction pipelines as evidenced in Intel Instruction Replay mechanism in its Itanium processor family \citep{bostian2013rachet}. \textbf{Data Privacy} – involving  transformations applied to legible data (plaintext user data) with the intent of making sure that it is only available to users that are authorized by the data owners. Typically, data encryption algorithms are used for achieving privacy guarantees, where the plaintext is transformed into cipher text before transmission and the keys needed for decrypting the encrypted cipher text are kept private. In order to secure web applications and systems, networks have to be able to promptly discern potential menaces and unwanted connections. Systems like intrusion detection (IDS) and intrusion prevention (IPS) are used for this purpose. Intrusion detection systems are divided in two groups: misuse detection (traditional IDS) and anomaly detection. Misuse detection systems are signature based, have high accuracy in detecting many kinds of known attacks but cannot detect unknown and emerging attacks. Our PTC, when properly trained according to previous knowledge of attacks, can be used as an intelligent comparator for the fast detection of misuse of the systems compared (performing convolution on string of data) to stored signatures of known exploits or attacks which are learnt in the photonic memories. This can be supplemented with anomaly-based intrusion detection as a prevention system. In fact, due to matrix multiplication and comparison performed at high-speed in optical accelerators, it can be used not just as smart pattern matcher, but as an evolving fast pre-screening of malicious activities, by collecting normal behaviors and detecting intrusion based on that, since new intrusion model can be implemented in the reprogrammable photonic memories thanks to the newly acquired and updating ‘knowledge’. Optical processing of high parallelism, inherent to several cryptographic operations, is enabled by taking advantage of various attributes of light waves such as the wavelength, phase, polarization, and amplitude. As such, energy-efficient, ultralow latency encryption-in-transit using optical accelerators can help address the grand challenges surrounding security in big data movement between computing systems. 

\subsection{Physics experiments}
\subsubsection{Intelligent Prefiltering in Astronomy and Scientific Applications}

\begin{figure}[h!]
\centering
\includegraphics[width=1\linewidth]{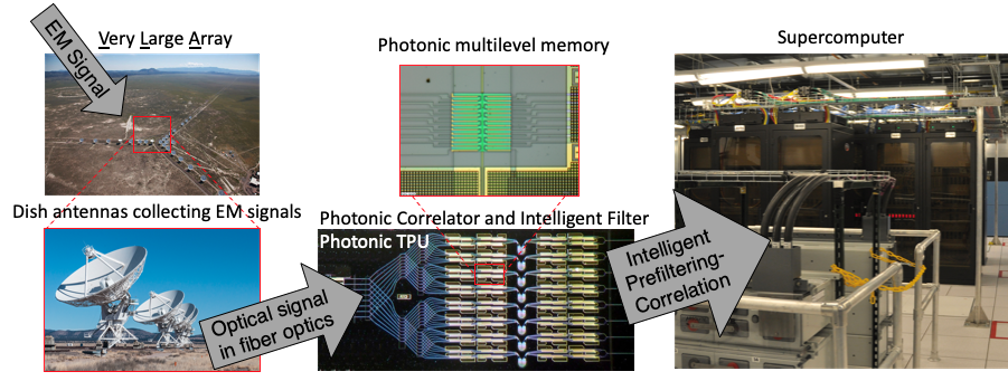}
% \vspace{-10pt}
\caption{Photonic tensor core and neural networks enable intelligent prefiltering and correlation for scientific discovery, such as between electromagnetic signals collected by Very-Large-Array telescope systems performing intelligent pre-filtering, thus reducing computation load in supercomputers. }
\label{fig:figure14}
\end{figure}

Astronomical radio observation and study on the galaxy formation have become extremely accurate thanks to the use of a plurality of telescopes arranged in an array, operating as a one single giant telescope. In this way, like all synthetic arrays, due to an enlarged equivalent aperture of 22 miles, the very large array (VLA), is sensitive and able to resolve a range of angular scales between the diffraction limit (Figure~\ref{fig:figure14}). A tremendous leap in this established technology is represented by the way the vast data obtained is collected and processed; data is fiber-optically fed to a supercomputer (WIDAR \citep{152noauthor_widar_nodate}) turning the VLA into a sensitive instrument. It is possible to obtain important information regarding star formation in 'interferometric' computing techniques, where the supercomputer correlates the hyper-spectral (wide band) signals from pairs of dishes obtaining a much sharper image than a single dish could produce. WIDAR uses FPGAs to perform correlations on the radiofrequency signal and only then sends relevant data to the cluster for further processing. Ultimately the cluster output is sent to an image-processing system. However, electronic data processing is limited by FPGA-setup times and fundamental electronic capacitive delay, resulting in delayed processing. To mitigate such processing limitations, photonic neural networks can be used as preprocessing unit to work synergistically with the WIDAR supercomputer on the vast data, in order to intelligently sorting and correlating the signal looking for specific chunks of radio-patterns (e.g.  hydrogen gas moves into galaxies to fuel star formation) in near real time ($\sim$ps). Besides the increased speeds and bandwidths that can come from working directly in the optical domain, leveraging on the intrinsic optical nature of signal captured by the dish-antennas travelling in optical fibers, the advantage of using the photonic architectures consists in exploiting the wave-nature of the input signals to perform inherent correlation detection or convolution using pre-stored/programmable trained weights, without active power consumption and burdensome electro-optic conversions, as discussed above. In this way, the total amount of information to be handled by the supercomputer and consequently by the cluster unit is reduced, saving resources for useful data, favored by intelligent pre-screening, towards resolving the evolution of the universe \citep{153noauthor_vla_nodate}. 

\subsubsection{Emerging applications: Quantum computer auxiliary systems and High-energy particle classification }

Scalable quantum computing depends on classical auxiliary technologies for state reconstruction, calibration, and control. Neural networks have been used for quantum state reconstruction \citep{154flurin2020using}, tomography \citep{155torlai2018neural}, and control \citep{156niu2019universal}. In some cases, the characteristic time constants of the state or instability are slow enough for a conventional computer to perform the task. In other cases, in particular for microwave qubits, the system is changing faster than a conventional computer can react. This means that a control and/or reconstruction task cannot occur in real-time. Photonic neural networks could reduce the latency of these operations, potentially opening up new opportunities to better monitor and stabilize quantum processing systems.

Photonic neural networks can exhibit latencies lower than electronic processors, whether neuromorphic, FPGAs, or ASICs. This low latency can make the crucial difference in certain applications where a decision must be made before the time to act passes. This critical time constraint exists in particle detectors such as CMS. Not all collisions are salient, so a trigger must classify the collision as salient or not before the next collision occurs. Hardware limitations dictate that the existing trigger uses rudimentary, non-adaptive algorithms that potentially overlook particular physics signatures. Recent work to improve the sophistication of particle classifications has adopted a neural network algorithms, implemented on FPGAs \citep{157duarte2018fast}. There is a potential for photonic neural networks to improve the performance of the time-critical triggering task, thus preserving physics signatures and enabling a higher collision rate.

\section{Conclusion} % (fold)
\label{sec:conclusion}

The emergence of neural network models has highlighted the importance of interconnects and parallel processing, which is inherently advantageous to implement in photonics. The research community has built bridges between photonic device physics and neural networks. The performance improvement of photonic neural networks is expected to continue as new devices (e.g., modulators or lasers) based on new materials and nanostructures demonstrate their potential of further increasing the efficiency. The next generation of photonic devices could consume only hundreds of aJs of energy per time slot, allowing analog photonic MAC-based processors to consume even less per operation~\citep{nahmias2019photonic,FerreiradeLimaTaitMehrabianNahmiasHuangPengMarquezMiscuglioElGhazawiSorgerShastriPrucnal+2020+4055+4073}. Meanwhile, advanced integration and fabrication techniques (e.g., silicon photonics) have provided an unprecedented platform to produce large-scale and low-cost photonic systems. The increased optical component density significantly extends the spectrum of information processing capabilities. Finally, monolithic fabrication, which integrates electronics and photonics on the same substrate, entails a tight co-integration of electronics and photonics, resulting in hybrid neuromorphic processors that can take the best advantages of both electronics and photonics depending on different applications.

In light of these developments, photonic neural networks have found places in many applications unreachable by conventional computing technology. Examples of applications explored in this paper include intelligent signal processing, high-performance computing, nonlinear programming, and control, enabling fundamental physics breakthroughs, etc. These applications particularly require low latency, high bandwidth, and low energies. To march ahead, we envisage a huge interest in developing the fundamental technologies (i.e., devices, fabrication/integration platforms, etc.) enabling large-scale photonic neural networks. In parallel, more applications will be identified and demonstrated, along with the photonic platform development, promising to expand the application space of AI and information processing.

\bibliographystyle{apacite}
\bibliography{reference}

\end{document}